\documentclass{aa}  

\usepackage{graphicx}
\usepackage{txfonts}
\usepackage{natbib}
\usepackage{float}
\usepackage[flushleft]{threeparttable}
\bibpunct{(}{)}{;}{a}{}{,} % to follow the A&A style
\usepackage[normalem]{ulem}

\usepackage{subfig}

\bibpunct{(}{)}{;}{a}{}{,} % to follow the A&A style
\usepackage[normalem]{ulem} % to follow the A&A style

\newcommand{\modif}[1]{\textnormal{#1}}
\newcommand{\modiff}[1]{\textnormal{#1}}
\newcommand{\modiflet}[1]{\textnormal{#1}}

\begin{document}

   \title{Transition disc nature of post-AGB binary systems confirmed by mid-infrared interferometry\thanks{Based on observations collected at the European Southern Observatory under ESO programmes 075.D-0605, 60.A-9275, 0104.D-0739, 105.20QN, 106.21NK, 109.23KF, and 110.24AC}}
   %Transition disc nature of post-AGB binary systems confirmed by mid-infrared interferometry
   %\hvw{the existence of} 

   \author{A. Corporaal \inst{1}
       \and  J. Kluska \inst{1}  
        \and H. Van Winckel \inst{1}
         \and K. Andrych \inst{2,3}
          \and N. Cuello \inst{4}
          \and D. Kamath \inst{2,3,5}
        \and A. M\'erand \inst{6}
           }

\institute{Institute of Astronomy, KU Leuven,
              Celestijnenlaan 200D, 3001 Leuven, Belgium\\
             \email{akke.corporaal@kuleuven.be}
             \and School of Mathematical and Physical Sciences, Macquarie University, Sydney, NSW, Australia
    \and Astronomy, Astrophysics and Astrophotonics Research Centre, Macquarie University, Sydney, NSW, Australia
    \and Univ. Grenoble Alpes, CNRS, IPAG, 38000 Grenoble, France
    \and INAF, Observatory of Rome, Via Frascati 33, 00077 Monte Porzio Catone (RM), Italy
    \and European Southern Observatory, Karl-Schwarzschild-Str. 2, 85748 Garching, Germany
    }
   \date{Received ; accepted}
   
\abstract 
   {Many properties of circumbinary discs around evolved post-asymptotic giant branch (post-AGB) binary systems are similar to those of protoplanetary discs around young stars.
   The deficits of near-infrared (near-IR) flux in the spectral energy distributions (SEDs) of these systems \modiflet{hints towards} large dust-free cavities that are reminiscent of transition discs as are commonly observed around young stars.}
   {We aim to assess the size of the inner rim of six post-AGB binary systems with lack in the near-IR like this. We used resolved mid-infrared (mid-IR) high-angular resolution observations of VLTI/MATISSE and VLTI/MIDI.
   The inner rim of only one such system was previously resolved.
   We compared these inner rim sizes to five systems with available MATISSE data that were identified to host a disc starting at the dust sublimation radius.}
   {We used geometric ring models to estimate the inner rim sizes, the relative flux contributions of the star, the ring, and an over-resolved emission, the orientation of the ring, and the spectral dependences of the components.}
   {We find that the inner dust rims of the targets with a lack of near-IR excess in their SEDs are $\sim2.5$ to 7.5 times larger than the theoretical dust sublimation radii, and inner rim sizes of the systems that do not show this deficit are similar to those of their theoretical dust sublimation radii.
   \modiff{The physical radii of the inner rims of these transition discs around post-AGB binaries are 3-25\,au, which are larger than the disc sizes inferred for transition discs around young stars with VLTI/MIDI.
   This is due to the higher stellar luminosities of post-AGB systems compared to young stars, implying larger dust sublimation radii and thus larger physical inner radii of the transition disc.}
   }
   %but smaller than the typical cavities inferred from millimetre interferometry.}
   {With mid-IR interferometric data, we directly confirm the transition disc nature of six circumbinary discs around post-AGB binary systems.
   Future observational and modelling efforts are needed to progress in our understanding of the structure, origin, and evolution of these transition discs.}
%5 {} token are mandatory

   \keywords{Stars: AGB and post-AGB - techniques: interferometric - binaries: general - protoplanetary disks - circumstellar matter}

   \maketitle
%
%-------------------------------------------------------------------
%Multi-wavelength infrared observations allow to study the inner disc regions in unprecedented detail. Constrains from geometric modelling of this data in our previous paper in this series put constrains on the disc inner rim structure as well as the radial structure. We used the three-dimensional Monte Carlo radiative transfer code MCMax3D to investigate the physical structure of the disc.
\section{Introduction}
%-protoplanetary discs, full discs and transition discs.
%- circumbinary discs around post-AGB binaries and description of the SED results and hypothesis.\\
%- link with depletion
%- AC Her \citep{Hillen_2015} + Anugu et al.\\
%- Infrared interferometry -- MATISSE observations to derive the cavity size\\
%- goal and outline\\

Numerous circumstellar discs with dust-free cavities have been detected and characterised around young stellar objects (YSOs). 
The properties of these transition discs around YSOs have recently been reviewed by \modif{\citet{vanderMarel_2023}}.
\modif{Transition discs are generally defined as circumstellar discs with a cavity in their inner dust distribution.
In these discs, dust is trapped at the edge of the cavity, which hinders the inward drift of dust grains.
Transition discs either have inner regions that are completely devoid of dust, or they have a small dust ring closer to the star, followed by a large cavity in the dust \citep[e.g. TW\,Hya;][]{Ratzka_2007, Menu_2014}.
}

The presence of \modif{transition} discs was first discovered based on a deficit of flux at near-infrared (near-IR) wavelengths in the spectral energy distributions (SEDs) \citep[e.g.][]{Calvet_2002}, and it was later confirmed by numerous millimetre interferometric observations with the Plateau de Bure Interferometer, the SubMillimeter Array, and the Atacama Large Millimeter/submillimeter Array (ALMA) \citep[e.g.][]{Pietu_2006,Brown_2009,Andrews_2011,Pinilla_2015}.

This lack of near-IR excess in the photometry has also been detected in circumstellar discs in a class of evolved binaries, namely in post-asymptotic giant branch (post-AGB) binaries. It indicates a large dust-free cavity similar to that of protoplanetary discs \citep[PPDs;][]{Kluska_2022}.
\modiflet{\citet{Kluska_2022}} provided a catalogue of the SEDs of Galactic post-AGB binary systems and classified the systems into five categories based on the IR excesses.
Systems with an excess starting in the near-IR are full discs with dust close to the dust sublimation radius, and they are classified as category 0 or 1 targets.
Systems of category 2, 3, and 4 lack this near-IR excess.
Category 2 systems show a strong mid-IR excess and are reminiscent of transition discs around YSOs.
Category 3 systems show a milder mid-IR excess, and these systems are likely a mix between a full disc and transition discs. In category 4 systems, finally, the IR excess starts later than the mid-IR. 
%while $\sim$10\% of the systems in the Galactic population are lacking such a near-IR excess and are reminiscent of transition discs around YSOs.
The properties of the circumbinary discs around post-AGB binaries are reviewed by \citet{VanWinckel_2018}.
These discs show Keplerian dynamics \citep{Bujarrabal_2013a, Bujarrabal_2013, Bujarrabal_2015, Bujarrabal_2017, Bujarrabal_2018, CallardoCava_2021, GallardoCava_2023} and a high degree of dust-grain processing \citep{Gielen_2011}, indicating a stable nature.

\citet{Kluska_2022} reported a relation between the IR excess colour, and thus the disc morphology, and a chemical phenomenon called depletion.
Depletion is a photospheric chemical anomaly with an underabundance of refractory elements (e.g. Fe or Ti) observed in the photosphere of the post-AGB primary \citep[e.g.][]{VanWinckel_1995, Giridhar_2005, Oomen_2018, Kamath_VanWinckel_2019}. It can be explained by a re-accretion of metal-poor gas from the circumbinary disc \citep[e.g.][]{Oomen_2019}. 
%\hvw{ Akke: also for full discs you need a trapping of the dust grains or at least a gas-dust separation and re-accretion of the gas}.
The needed condition is that the dust grains are trapped within the disc, while the gas, devoid of refractory elements, can be accreted. 
The efficiency of this gas-dust separation is linked to the disc structure. %\JK{'depends' --> 'is linked to' (as 'depends' implies that there is a causal relation)}

%While this idea could explain the depletion pattern of systems surrounded by a full disc, in which the dust grains feel the radiation pressure, it cannot explain the strong depletion observed in systems with a dust-free cavity.
%Such a link points towards accretion processes and the gas-dust separation depends on the disc structure.
The relation is such that depletion is stronger for objects with a lack of near-IR excess, and \citet{Kluska_2022} proposed that the depletion process is linked to the process creating the cavity.
These authors suggested that this gas-dust separation might arise because a giant planet traps the dust in the outer disc and in that way carves a cavity, following \citet{Kama_2015}, or it might arise by dynamical truncation of the binary \citep{Artymowicz_1994, Miranda_2015,Hirsh_2020} .

To progress on this hypothesis, we need to spatially resolve the inner rim regions, which are accessible with IR interferometric observations.
These observations of systems that are surrounded by a full disc show that dust is located close to the dust sublimation radius and extends outwards \citep{Hillen_2016, Hillen_2017, Kluska_2019, Corporaal_2021}. 
In contrast, mid-IR interferometric observations with the MID-infrared Interferometric instrument (MIDI), a decommissioned two-beam combiner on the Very Large Telescope Interferometer (VLTI), of AC\,Her (a category 2 system) have resolved the inner dust cavity, which is about seven times larger than its theoretical dust sublimation radius \citep{Hillen_2015}.
AC\,Her is the only post-AGB binary system for with a large dust-free cavity was resolved so far.
Near-IR observations of AC\,Her with the long-baseline interferometric array at the Center for High Angular Resolution reveal the three-dimensional orbit of the inner binary and show that the inner cavity cannot be explained by the dynamical truncation of the binary \citep{Anugu_2023}.

Here we present mid-IR interferometric observations with the four-beam combiner Multi-AperTure mid-Infrared SpectroScopic Experiment (MATISSE) installed on the VLTI of ten Galactic post-AGB binary systems together with the reanalysis of the archival MIDI data of AC\,Her.
We aim to constrain the sizes of the cavities of the transition disc candidates and compare them to full-disc systems using geometric models.
We present our sample selection and observations in Sect.\,\ref{sect:observations}, describe our geometric modelling of these data in Sect.\,\ref{sect:geometricmodelling}, and analyse the results in Sect.\,\ref{sect:results}.
We discuss the inferred cavity sizes in Sect.\,\ref{sect:discussion} and summarise our conclusions in Sect.\,\ref{sect:conclusion}.

%\JK{I would directly start with protoplanetary disks and transition disks and then going to post-AGB disks in which we also suspect transition disks from photometry.
%Then link to depletion
%Then AC Her as the only measurement of a large cavity and then introduce the MATISSE observations to confirm cavity sizes around other post-AGB binaries.}

\section{Observations}
\label{sect:observations}
The SEDs were taken from the catalogue of Galactic post-AGB binary systems of \citet{Kluska_2022}. 
We used mid-IR interferometric observations of post-AGB binaries obtained with VLTI/MATISSE \citep{Lopez_2022}.
MATISSE operates in the $L$ (2.9-4.2\,$\mu$m), $M$ (4.2-5.0\,$\mu$m), and $N$ (8-13\,$\mu$m) bands.
MATISSE combines the light of either the four Unit Telescopes (UTs) or the four Auxiliary Telescopes (ATs) and provides six simultaneous baselines and three independent closure phases (CPs) per measurement.
The interferometric observables are the visibilities and the CPs.
The visibility is measured as a squared visibility amplitude ($V^2$) in the $L$ band and the visibility amplitude ($|V|$) or the correlated flux (CF) in the $N$ band.

Our data set contains five systems of category 1 (HD\,101584, HD\,109014, IW\,Car, IRAS\,08544-4431 (hereafter IRAS\,08544), and IRAS\,15469-5311 (hereafter IRAS\,15469)) with an IR excess starting at near-IR wavelengths, and six systems (CT\,Ori, EP\,Lyr, AC\,Her, AD\,Aql, RU\,Cen, and ST\,Pup) with an excess starting at mid-IR wavelengths (categories 2 and 3).
The target selection for the transition disc candidates was based on (1) the lack of near-IR excess, (2) the characteristics in depletion tracers ([Fe/H] $\sim -2.0$, [Zn/Ti] $\sim 2.0$, and $T_\mathrm{turn-off} \sim 1000$\,K; see also Table \ref{table:stellarprops} and \citet{Kluska_2022}), and (3) the observability of the systems with MATISSE. 
%\JK{It would have been nice to introduce the categories to more easily justify (1) and (2). Maybe already in the intro?} 
%The MATISSE observations have been taken for five out of \hvw{mmm i don't get the numbers here: 11 sources in total, not 8 no?. What is meand by five out of eight???} eight of these systems so far.
The transition disc sample was complemented with data of AC\,Her taken with the VLTI/MIDI, which operated in the $N$ band as it is the only system for which a dust-free cavity was detected so far. 
For a log of these observations of AC\,Her, we refer to \citet{Hillen_2015}.
The data of the transition disc candidates were compared to full-disc systems in the MATISSE wavelength range.

All MATISSE observations were taken in hybrid mode with a low spectral resolution ($R\sim30$) between 2019 and 2023.
For a log of the observations of IRAS\,08544, we refer to \citet{Corporaal_2021}.
Data of IRAS\,15469 were taken with the ATs in the small configuration (program ID 105.20QN, PI: Kluska).
Observations of IW\,Car were carried out using the ATs on the small and medium configurations (program ID 106.21NK, PI: Kluska).
Data of HD\,101584, HD\,108015, EP\,Lyr, AD\,Aql, RU\,Cen, CT\,Ori, and ST\,Pup were taken using the UTs with the GRA4MAT mode using GRAVITY as a fringe tracker to stabilise the fringes of these faint targets (program IDs 109.23KF and 110.24AC, PI: Corporaal).
The log of these observations is reported in Table \ref{tab:MATISSE_obs}.

The data were reduced and calibrated with the MATISSE data reduction pipeline \modif{version 1.7.5}.
We used all data for which no observational or calibration problems occurred (see Appendix\,\ref{sect:datareduction}).
The $N$-band data taken with the UTs were reduced in CF mode, meaning that photometry was not taken during the observations, and therefore, the coherent flux measurements could not be normalised. 
These visibilities are therefore reported in CF.

Calibrated visibilities and the corresponding $uv$ coverages are shown in Figs. \ref{fig:model_data_Lband} and \ref{fig:model_data_Nband}.
\modif{All systems are well resolved. 
HD\,108015, AD\,Aql, RU\,Cen, and ST\,Pup show a prominent silicate feature at 11.3\,$\mu$m in their $N$-band visibilities.
}

%\JK{Target selection and quick description on what the data looks like. V2, CF, CP}

\section{Geometric modelling}
\label{sect:geometricmodelling}
%- PMOIRED \citep{Merand_2022} \\
%- Explanation of the modelling strategy; the components, when we fit what, first zero and size\\
%- Dust sublimation radius: formula, ranges (i.e. explanation of 1000\,K, 1250\,K and 1500\,K)
\begin{figure}
\centering
  \resizebox{\hsize}{!}{\includegraphics[width  =12cm]{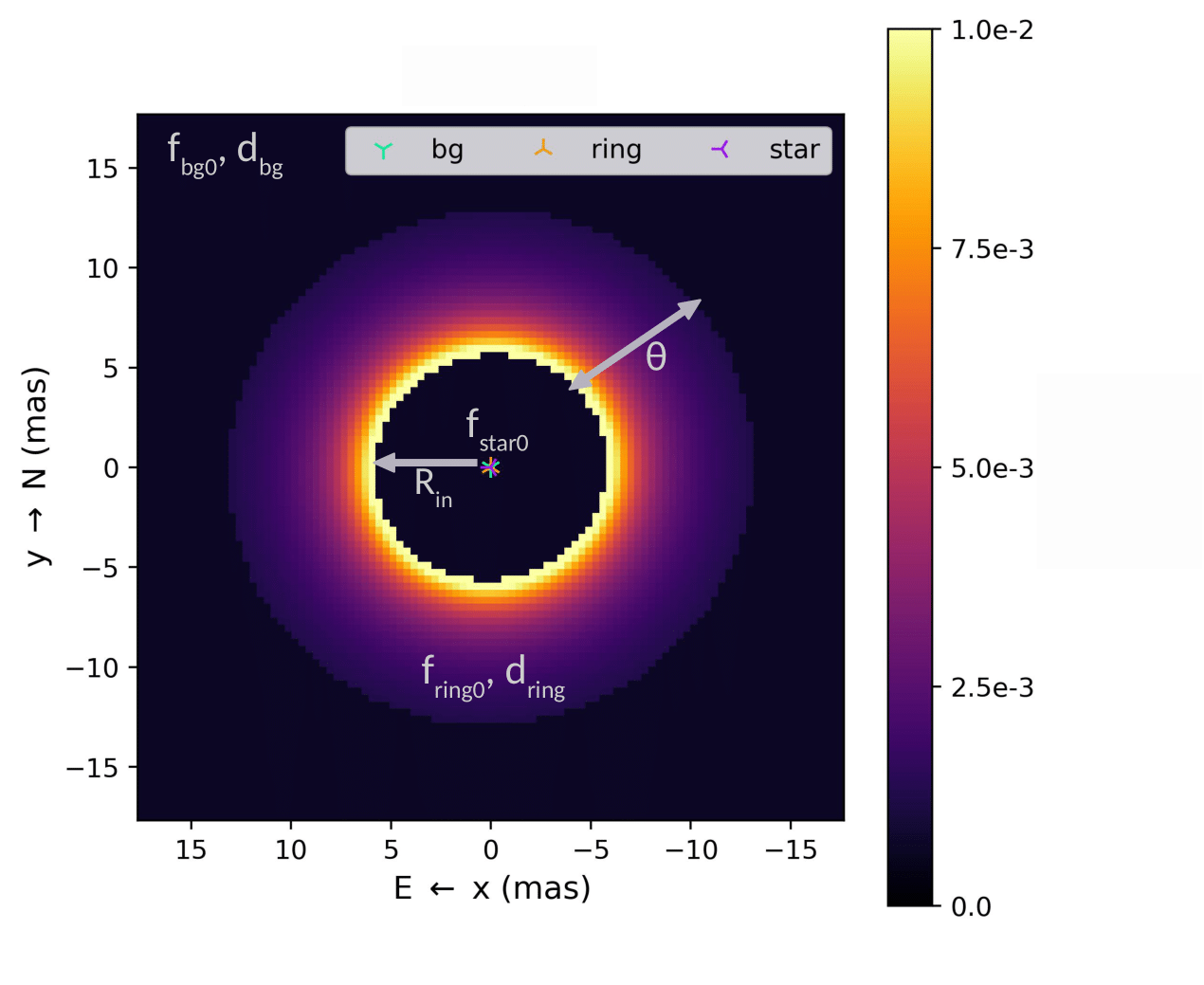}}
  \caption{Illustration of the disc structure including a central star, an inclined ring, and an over-resolved component (bg).
  \modif{This illustration shows the best-fit model image of IRAS\,08544 in the $L$ band: an almost pole-on disc with an inclination of $\sim 16$\,deg.}
  The colour bar indicates the surface density in arbitrary units. 
  The inner rim is brighter, and the surface density decreases \modif{with $R^{-3}$ ($p=-3$).}
  }
 
  \label{fig:structure_disc}
\end{figure}

\begin{figure*}
    \centering
    \includegraphics[width=17cm]{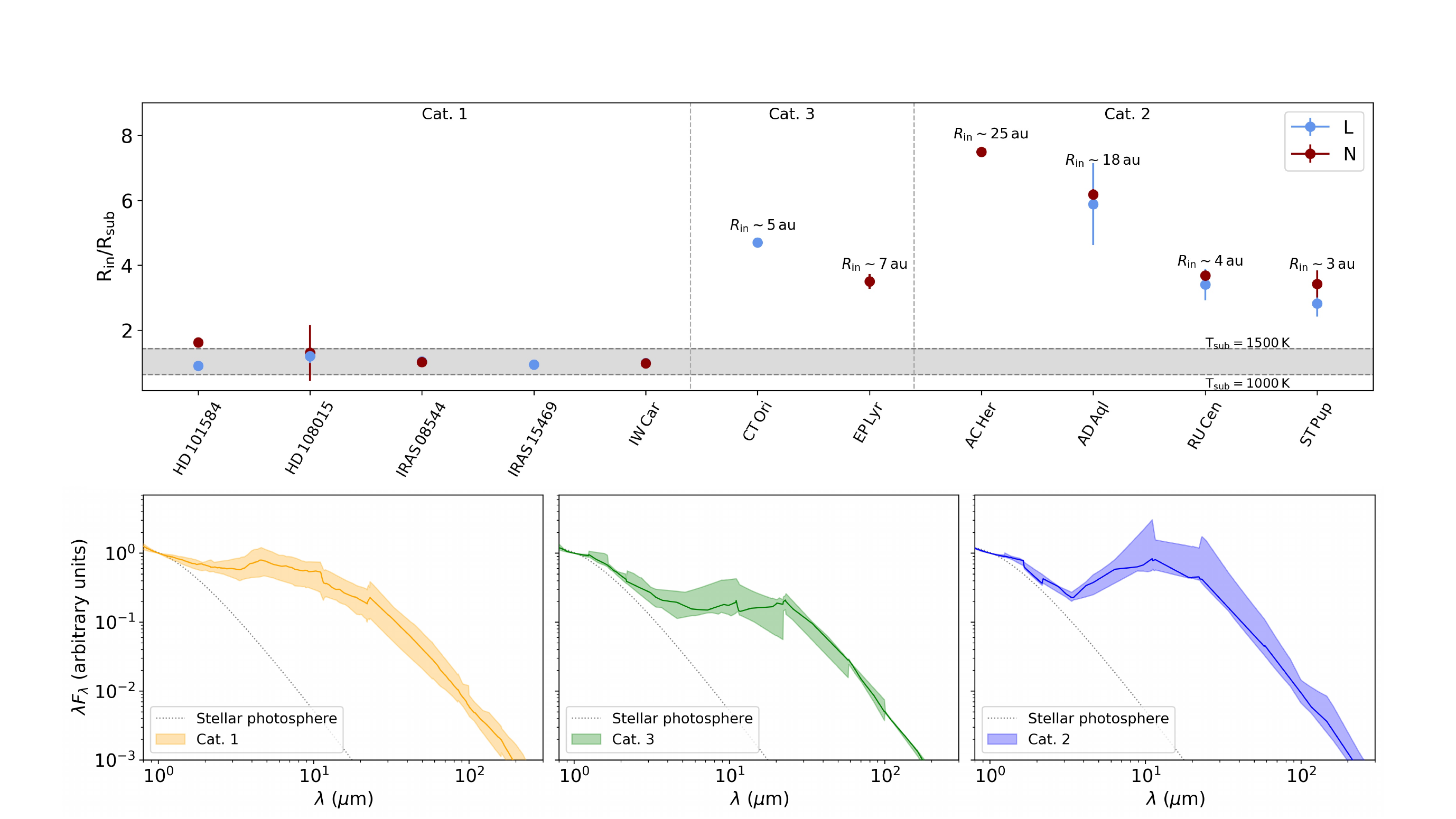}
    \caption{{\modiff{Cavity size diagram for the 11 systems (\textit{top})} and the corresponding median SEDs (\textit{bottom})}. 
    \modiff{The cavity sizes are displayed relative to the theoretical dust sublimation radii.}
    The shaded grey area corresponds to the dust sublimation radii with temperatures of 1000-1500\,K.
    The physical sizes of the inner rim radii are indicated for the transition discs assuming \textit{Gaia} DR3 distances. 
    The categories correspond to the classification of the SEDs by \citet{Kluska_2022}.
    The median SEDs of sources in categories 1, 3, and 2 are shown below and indicate the differences in the IR excesses between each category.
    }
    \label{fig:rinrsub}
\end{figure*}

To constrain the inner rim radii, we modelled the IR interferometric visibilities using the code called parametric modeling of optical interferometric data \citep[PMOIRED\footnote{https://github.com/amerand/PMOIRED};][]{Merand_2022}.
PMOIRED is a geometric modelling tool for IR interferometry that relies on semi-analytical expressions of complex visibilities.
PMOIRED allows modelling systems by linearly adding the expressions of the considered components.

The systems were considered to consist of either a single star, an inclined ring, and an over-resolved emission or a binary and an over-resolved emission.
The choice between models was based on the SEDs of the targets.
In all but one target, the IR excess starts at wavelengths between 1\,$\mu$m and 4\,$\mu$m, indicating that the circumbinary disc emits in the $L$ and $N$ bands and that we thus probe the disc in both bands.
In EP\,Lyr the IR excess starts later than $3-4\,\mu$m, but before $\lambda = 8\,\mu$m, such that the system is dominated by stellar emission in $L,$ but the disc emits in $N$.
%EP\,Lyr is a confirmed binary system \citep{Manick_2017}.

\modif{Fig. \ref{fig:structure_disc} illustrates the presumed structure of the ring models.}
\modiff{In our modelling, we aimed to constrain the inner rim sizes and therefore did not consider a dust distribution with a small dust ring closer to the star followed by a cavity.}
In all disc models, we neglected the illumination of the companion, although we expect all systems to be binaries \citep[e.g.][]{VanWinckel_2018}.
The central star was considered in all cases where its relative flux contribution as deduced from the SED was $>2\%$.
The flux contributions of the components, $f_{i}$, were defined at the central wavelength, $\lambda_0$, of each band such that $\Sigma f_i(\lambda = \lambda_0) =1$.
The central star was modelled as a point source with a spectrum approximated by a Rayleigh–Jeans tail proportional to $\lambda^{-4}$ in the $L$ and $N$ bands ($T_\mathrm{eff} = 5000-7500$\,K).
The over-resolved emission was modelled as a wavelength-dependent fully resolved component to account for all large-scale scattered light emission with a spectrum following
\begin{equation}
    \label{eq:discspectrum}
    f_\mathrm{bg} = f_\mathrm{bg_0}\left(\frac{\lambda}{\lambda_0}\right)^{d_\mathrm{bg}},
\end{equation}
where $d_\mathrm{bg} = \frac{d \log F_\lambda}{d \log\lambda}$ is the spectral index of this over-resolved emission.
Visibility data displayed in CF are not sensitive to the over-resolved emission.
Instead of $d_\mathrm{bg}$ and $f_\mathrm{bg_0}$, the total flux at zeroth baseline, $F_\mathrm{tot}$, was fitted for these data.
The orientation of the circumbinary ring is given by its inclination, $i$, and position angle, $PA$, measured north to east.
The ring was assumed to be continuous with a radial profile $\propto R^p$ and to have a wavelength-dependent spectrum following Eq. \ref{eq:discspectrum}.
The angular emission of the ring was parametrised by the inner rim radius, $R_\mathrm{in}$, and the ring width, $\theta$.
This resulted in up to nine free parameters.
\begin{table*}[t]
\caption{Best-fit parameters of the geometric models in the $L$ band}
\label{table:bestfit_L}

\centering
\begin{threeparttable}
\begin{tabular}{lcccccccccc}
\hline
&&&&&Ring models
 \\
\hline
Object &   $\chi^2_\mathrm{red}$ &       $R_\mathrm{in}$ &    $\theta$  &   $i$&        PA  & $p$ & $f_\mathrm{bg_0}$ &$f_\mathrm{star_0}$& $d_\mathrm{bg}$ &  $d_\mathrm{ring}$  \\
& &  (mas) & (mas)& (deg) & (deg) & & (\%)& (\%) \\
  \hline

HD\,101584 & 0.7 & 1.9$\pm$0.1 & 6.5$\pm$0.6 & 23$\pm$17 & 18$\pm$21 &-3.0$\pm$0.3 & 11$\pm$9 & 2  &  34.1$\pm$10.0 &  24.9$\pm$6.8\\
%HD\,108015 & \modif{15} & \modif{1.7$\pm$0.1} & \modif{8.1$\pm$1.7} & \modif{42$\pm$10} & \modif{93$\pm$12} &-1.4 & \modif{5$\pm$1} & 10 &  \modif{-15.4$\pm$3.7} &   \modif{`4.2$\pm$3.8} \\
\modif{HD\,108015$^{*}$} & \modif{9.0} & \modif{1.7$\pm$0.3} & \modif{8.1$\pm$1.7} & \modif{35$\pm$11} & \modif{45$\pm$13} &-1.4 & \modif{5$\pm$1} & 10 &  \modif{-15.4$\pm$3.7} &   \modif{4.2$\pm$3.8} \\
IRAS\,08544 & 23 &5.8$\pm$0.1 & 7.9$\pm$0.9 & 16$\pm$5&   116$\pm$25 & -1.6 & 37$\pm$1 & 6 & 2.8$\pm$1.4 &  -0.8$\pm$0.8\\
IRAS\,15469 & 1.0 & 1.7$\pm$0.2 & 14.8$\pm$2.3 & 47$\pm$5 &     50$\pm$4 &  -2.1 & 11$\pm$3 & 6  &   -1.1$\pm$5.3 &  -1.5$\pm$5.8\\
\modif{IW\,Car$^{*}$} &\modif{29}& \modif{2.2$\pm$0.2} & \modif{18.2$\pm$1.1} & \modif{50$\pm$3} &    \modif{172$\pm$2} & \modif{-1.1$\pm$0.1} & \modif{11$\pm$1} & \modif{2}  &   \modif{34.2$\pm$5.4} &  \modif{34.9$\pm$5.3}\\
%CT\,Ori & \modif{7.1} & \modif{3.6$\pm$0.8} & \modif{2.1$\pm$1.1} & \modif{48$\pm$24} & \modif{205$\pm$23}  &\modif{-1.4$\pm$0.1} & \modif{18$\pm$3} & \modif{29} &    \modif{2.0$\pm$1.6} &  \modif{-1.5$\pm$0.9}\\
\modif{CT\,Ori$^{*}$} & \modif{1.5} & \modif{3.9$\pm$0.2} & \modif{2.1$\pm$1.0} & \modif{40$\pm$4} & \modif{172$\pm$5}  &\modif{-1.5$\pm$0.3} & \modif{20$\pm$3} & \modif{29} &    \modif{2.9$\pm$1.8} &  \modif{-0.4$\pm$1.8}\\
AD\,Aql& 0.5 & 2.2$\pm$0.5 & 2.8$\pm$1.3 & 39 & 48$\pm$19 &-2 & 27$\pm$11 & 37 &  8.3$\pm$2.1 &  12.5$\pm$3.7 \\ 
RU\,Cen  & 1.6 & 3.1$\pm$0.4 & 0.9$\pm$0.5 & 68$\pm$8 &    154$\pm$5 &  -2 &9$\pm$2 & 78 &  9.2$\pm$1.6 &  -6.3$\pm$3.1 \\
ST\,Pup & 15 & 1.5$\pm$0.6 & 9.8 & 50 & 128 & -1.6 &30$\pm$5 & 61 &  6.2$\pm$1.0 &  15$\pm$12 \\ 
\hline
&&&&&Binary models
\\
\hline
Object &   $\chi^2_\mathrm{red}$ & $f_\mathrm{bg_0}$&  $f_\mathrm{star}$& $f_\mathrm{star2}$ &  $d_\mathrm{bg}$ &  $x_\mathrm{star2}$ & $y_\mathrm{star2}$ \\
& &(\%)  &(\%)&(\%)&&(mas) & (mas)\\
  \hline
EP\,Lyr & 2.3  &  18$\pm$17 & 65$\pm$12 & 17$\pm$14 &  2.3$\pm$2.1 &  -0.1$\pm$ 0.3 &  1.6$\pm$0.8\\
\hline
\\
\end{tabular}
\end{threeparttable}
\tablefoot{Reported values are calculated at $\lambda_0=3.5\,\mu$m.

\modif{$^*$ The model included azimuthal modulations.}}

\end{table*}

\begin{table*}[t]
\caption{Best-fit parameters of the geometric models in the $N$ band}
\label{table:bestfit_N}
\centering
\begin{threeparttable}
\begin{tabular}{lccccccccccc}

 \\
\hline
Object &   $\chi^2_\mathrm{red}$ &       $R_\mathrm{in}$ &    $\theta$  &   $i$&        PA & $p$ & $f_\mathrm{bg}$ &$f_\mathrm{star_0}$&  $d_\mathrm{bg}$ &  $d_\mathrm{ring}$& $F_\mathrm{tot}$\\
& &  (mas) & (mas)& (deg) & (deg)&&(\%)&(\%)&&& (Jy)\\
  \hline

HD\,101584 & 4.7& 2.6$\pm$0.1 & 15.5$\pm$0.3 & 23 & 18 & -1.1$\pm$0.1 & - & - & - & -1.3$\pm$0.2 & 63.4$\pm$0.4 \\
HD\,108015 & 1.1 & 1.8$\pm$0.1 & 12.5$\pm$0.3 & 32$\pm$2 & 15$\pm$2 &-1.3$\pm$0.1 &-& -& -&-0.2$\pm$0.1 & 23.7$\pm$0.3 \\
IRAS\,08544 & 1.8 & 6.1$\pm$0.3 & 37.1$\pm$0.2 & 16 & 45$\pm$4 &-1.6$\pm$0.1 & & - & 1.7 & -1.4$\pm$0.1 & 177$\pm$3 \\
IW\,Car& \modif{4.4} & \modif{2.4$\pm$0.1} & \modif{27.8$\pm$0.4} & \modif{64$\pm2$}&\modif{174$\pm 1$} &\modif{-1.2$\pm$0.1} & \modif{30$\pm$1} & - & \modif{-14.8$\pm$10.5} & \modif{-19.7$\pm$10.3} & -\\
EP\,Lyr & 1.2 & 3.1$\pm$0.3 & 43.9$\pm$1.6 & 64$\pm$2 & 171$\pm$3 & -1.9$\pm$0.1 &-& 1 &-& -2.6$\pm$0.1 & 0.16$\pm$ 0.01 \\
AC\,Her & 9.7 & 18.0$\pm$0.3 & 30.9$\pm$0.5 & 50 & 166$\pm$2 & -2& 6$\pm$1 & 2  &7.1$\pm$1.2 & 10.0$\pm$2.1  & -\\
AD\,Aql & 7.2 & 2.3$\pm$0.1 & 3.6$\pm$0.2& 39$\pm$15 & 66$\pm$11 & -2& - & - & - & 0.8$\pm$ 0.1 &1.2$\pm$0.1\\ 
RU\,Cen  & 50 & 3.4$\pm$0.1 & 32.8$\pm$0.7 & 68$\pm$1 & 166$\pm$4 & -1.7 & -& -& - & 1.3$\pm$0.1 & 5.4$\pm$0.2\\
ST\,Pup &  84 & 1.8$\pm$0.3 & 29.5$\pm$2.4 & 50$\pm$2 & 128$\pm$2 & -1.6$\pm$0.1 & -&-&-&0.5$\pm$0.2 & 5.1$\pm$0.4\\

  \hline
\end{tabular}
\end{threeparttable}
\tablefoot{Reported values are calculated at $\lambda_0=10.5\,\mu$m.}

\end{table*}

The binary model took into account the flux ratios of the primary star, the secondary star (both modelled as point sources), and the over-resolved component.
The positions of the secondary with respect to the primary, $x_\mathrm{star2}$ and $y_\mathrm{star2}$, were allowed to be free.
This resulted in five independent free parameters.

Since $R_\mathrm{in}$, $f_\mathrm{star}$, and $\theta$ are correlated with each other, we chose to let them free one by one. 
When the relative flux of the star was predicted to be similar to the flux ratio as deduced from the SED, we fixed it to this value and attempted to constrain the other free parameters.
Then we explored each parameter one by one until all parameters that could be constrained with the available data had reasonable estimates.
In each step, we \modif{minimised} the reduced $\chi^2$, $\chi^2_\mathrm{red}$.
For each target, we first attempted to fit the $L$- and $N$\modif{-}band data independently, but for some targets, the data did not allow us to constrain some parameters.
In these cases, we fixed these parameters to the best-fit values of the other band or to the values fitted in an intermediate step.
These parameters are reported without uncertainties.
\modiflet{The uncertainties were calculated using a data resampling (bootstrapping) algorithm available in PMOIRED and scaled to the data scatter.}

Our main aim was to constrain the inner rim size, which is obtained from the location of the fist zero of the visibilities. 
We therefore focused on reproducing this first zero.
We checked whether the fitted $R_\mathrm{in}$ was not strongly correlated with other fitting parameters.
To have a distance-independent measure, the inner rim size was then compared to the theoretical dust sublimation radius, $R_\mathrm{sub}$ \citep{Monnier_2002, Lazareff_2017},
\begin{equation}
\label{eq:Rsub}
    R_\mathrm{sub} = \frac{1}{2}\left(\frac{C_w}{\varepsilon} \right)^{1/2} \left(\frac{L_\mathrm{bol}}{4\pi\sigma T_\mathrm{sub}^4}\right)^{1/2},
\end{equation}
where $C_w$ is the backwarming coefficient, $\varepsilon$ is the cooling efficiency, $L_\mathrm{bol}$ is the bolometric luminosity of the post-AGB star, $\sigma$ is the Stefan-Boltzmann constant, and $T_\mathrm{sub}$ is the dust sublimation temperature.
We assumed $C_w \sim 1$ and $\varepsilon \sim 1$.
\citet{Kluska_2019} found that for post-AGB discs, the temperature of the inner dust rim is $\sim 1200-1300$\,K.
Similarly, radiative transfer modelling of IRAS\,08544 shows that temperature of the inner dust rim is $\sim 1250$\,K \citep{Corporaal_2023}.
Therefore, we calculated $R_\mathrm{sub}$ at $T_\mathrm{sub}=1250$\,K.

As a consequence of the focus on reproducing the first zero, larger baseline data and hence smaller angular scales were not always fitted well. 
For these systems, we imposed a more complex model in which we added an azimuthal modulating in the ring.
This modulation captures variations in the intensity distribution of the inner rim.
We used the first- and second-order azimuthal harmonics implemented in PMOIRED as sinusoidal functions.
the fitted parameters were the first- and second-order amplitudes of these modulations and the corresponding position angles with respect to the PA of the ring.

%\LEt{a single sentence does not constitute a paragraph. Please either add to this or merge}.

%We started the fitting of each object in each band by setting the initial guesses of each parameter.
%Initial guesses of the full discs were taken from \citet{Kluska_2019} and \citet{Corporaal_2021}.
%The contribution of the star relative to the total flux was set based on the SED.
%The initial guess of the flux at zero baseline, $F_\mathrm{tot}$, was set to the value of the correlated flux at the shortest baseline.

\begin{figure}
\centering
  \resizebox{\hsize}{!}{\includegraphics[width=12cm]{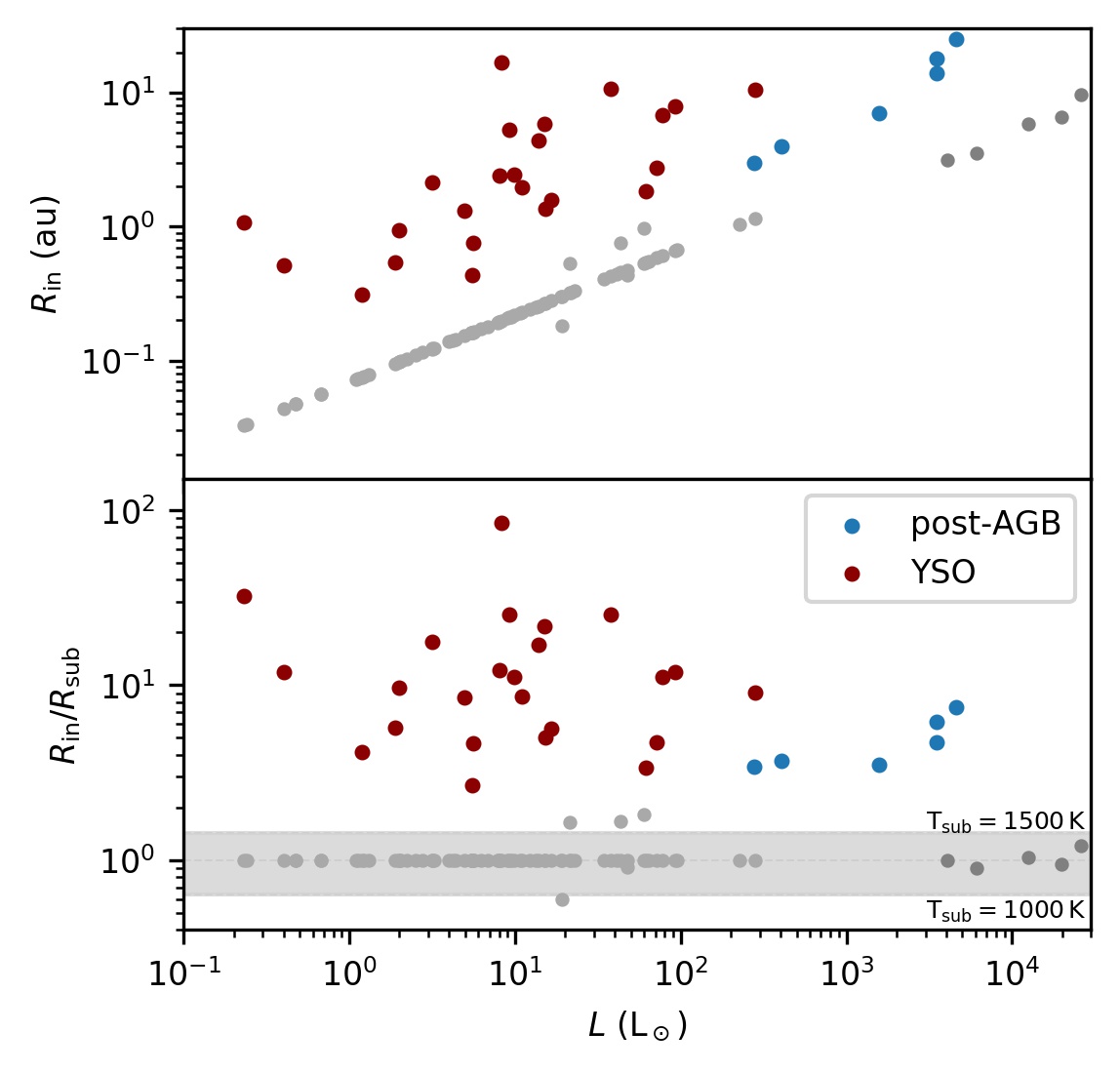}}
  \caption{Comparison between the dust-free cavity sizes as revealed with mid-IR interferometry of the six post-AGB binary systems hosting transition discs in our sample and transition discs around YSOs from \citet{Varga_2018} as a function of stellar luminosity.
  The sizes are displayed as either the physical size (\textit{top}) or as the inner rim size relative to the dust sublimation radius (\textit{bottom}). 
  The latter is independent of the distance uncertainties.
  The sample of full discs of both the YSOs and post-AGB discs are overlaid in light grey \citep[from][]{Varga_2018} and dark grey (this work).
  %The observations were taken with similar angular resolutions of $\sim 11$\,mas.
 }
  \label{fig:comp_ysos}
\end{figure} 

%Since the $uv$ coverage is sparse we cannot constrain all parameters based on single observations. 
\section{Results}

\label{sect:results}
%- Figure of full discs vs transition discs sizes as a function of their disc sublimation radius\\
%- Disc temperature comparison?\\
% - Something with the projected separation? This can be the right axis of the figure
The parameters of the best-fit geometric models that reproduce the visibility data of the targets in the $L$ and $N$ bands are displayed in Tables \ref{table:bestfit_L} and \ref{table:bestfit_N}, respectively.
The visibility curves of the best-fitting models in the $L$ and $N$ bands are shown in Fig. \ref{fig:model_fits_Lband} and Fig. \ref{fig:model_fits_Nband}, respectively.
The inner rim radii relative to the theoretical dust sublimation radii of the targets are displayed in Fig. \ref{fig:rinrsub}.

The inner rim radii of the category 1 targets are well constrained and are all within the boundaries of dust sublimation temperatures of 1000-1500\,K, at least in $L$. 
The inner rim radii of category 2 and 3 targets are clearly larger and are 2.5 to 7.5 times the dust sublimation radius. 
This translates into dust-free cavities of $\sim 3-25$\,au assuming \textit{Gaia} DR3 distances \citep{Gaia_2016, Gaia_DR3_2022}.
In all cases for which reliable $L$\modif{-} and $N$\modif{-}band data were obtained, the estimates of the inner rim radii are consistent between these bands.

IW\,Car and EP\,Lyr are RVb variables \citep[i.e. they show long-term changes in the mean brightness; e.g.][]{Kiss_2007, Manick_2019} \modif{and indeed show high inclinations ($\sim60^\circ$) in the $N$ band.
The disc of EP\,Lyr does not emit in $L,$ and thus no inclination can be derived.
The inclination derived for IW\,Car in $L$, however, is strongly preferred to be lower ($i\sim50$\,deg) than that of other RVb stars (see also Sect.\,\ref{sect:discussion}).}

The radial intensity profiles are better constrained in $N$, covering broader disc regions than in $L$.
However, this is specifically the case for the full-disc sources, where MATISSE probes beyond the inner rim.
\modif{Some $\chi^2_\mathrm{red}$ are relatively high.} 
We note that not the entire complexity has been captured by our modelling.
\modif{In $L$, higher $\chi^2_\mathrm{red}$ values are found for the systems with a substantial amount of data (IRAS\,08544 and IW\,Car), for the system to which the disc does not contribute significantly (9\% at 3.5\,$\mu$m, ST\,Pup), or for the systems in which the complexity at smaller scales is not fully fitted (HD\,108015 and CT\,Ori).}

\modif{
To reproduce the larger baselines of IW\,Car, HD\,108015, and CT\,Ori azimuthal modulations of the first- and second-order were taken into account.
These modulations lower the $\chi^2_\mathrm{red}$ minimisation results from 60 to 29 for IW\,Car, from 18 to 9.0 for HD\,108015, and from 7.1 to 1.5 for CT\,Ori.
The best-fit results of IW\,Car have variations with amplitudes of 0.52$\pm$0.13 and 0.35$\pm$0.05 and position angles of -79$\pm$24\,deg and -48$\pm$6\,deg with respect to the $PA$ of the ring for the first- and second-order modulation, respectively.
Similarly, the best-fit results of HD\,108015 have azimuthal modulations with amplitudes of 0.68$\pm$0.16 and 0.51$\pm$0.15 at position angles of 122$\pm$13\,deg and 82$\pm$10\,deg for the first- and second-order modulation, respectively.
Despite this additional complexity, the model is not able to fully reproduce the data of the large baseline, which corresponds to baseline UT1-UT4, while it is able to reproduce the data of the other baselines.
The best-fit results of CT\,Ori show azimuthal modulations with amplitudes of 1.0$\pm$0.1 and 0.25$\pm$0.21 with position angles of 90$\pm$11\,deg and -87$\pm$23\,deg for the first- and second-order modulation, respectively.
}

\modif{
The large baseline data of ST\,Pup are not well reproduced.
As a result of the small contribution of the disc to the total flux, any azimuthal modulations of the inner rim do not improve the fit of these baselines.
The contribution of its over-resolved component is high.
Smaller-scale structures such as the binary component and their respective offsets from the centre likely play a role, but have not been captured in our modelling.}

\modif{
In $N$, higher $\chi^2_\mathrm{red}$ values are for systems that show prominent silicate features in their visibility curves (AD\,Aql, RU\,Cen, and ST\,Pup) or for a system with a sparse $uv$ coverage (AC\,Her).
For AD\,Aql, RU\,Cen, and ST\,Pup, only the general trend could be reproduced with these geometric modelling because no specific dust species was assumed.
The small-baseline data (probing larger scales) of AC\,Her are not well reproduced with our modelling, while the larger-baseline data and the location of the first zero are reproduced well.
}

\section{Discussion}
\label{sect:discussion}
%- Comparison with YSOs and discussion on stated hypothesis on the giant planet?
 %- Link with depletion? But inner cavity size related to the depletion patterns?
The inner rim radii of the transition disc candidates are all several times larger than the dust sublimation radius. This indicates that these systems indeed show large inner dust-free cavities. 
CT\,Ori, EP\,Lyr, AC\,Her, AD\,Aql, RU\,Cen, and ST\,Pup are therefore indeed transition discs, reminiscent of those that surround YSOs.
The physical inner cavity sizes are between 3 and 25\,au. 
For AC\,Her, we found a $\sim 10\%$ lower $R_\mathrm{in}$ than inferred with the radiative transfer model of \citet{Hillen_2015} (scaled to the same distance) because we focused on reproducing the first zero.

\modif{
The inclination derived for IW\,Car in $N$ is consistent with the RVb phenomenon.
However, the $L$-band visibilities prefer lower inclinations of $\sim50$ deg.
Similar inclinations were found with near-IR interferometric observations by \citet{Kluska_2019}.
While azimuthal modulations significantly improve the fit in the $L$ band, the $\chi^2_\mathrm{red}$ remains high, and our simple geometric model falls short of capturing the complexity of the disc structure. 
Multi-wavelength analyses combining the near-IR and mid-IR observations are needed to reveal its complexity.}

\modif{
For four systems, a silicate feature is evident in the $N$-band visibility data.  
The most prominent silicate features are seen in the systems hosting a transition disc.
Only in one system hosting a full disc, HD\,108015, is the silicate feature observed, but it is much weaker than for AD\,Aql, RU\,Cen, and ST\,Pup. 
A radiative transfer treatment with the inclusion of dedicated dust species is needed to characterise this feature.}

\subsection{\modif{Comparisons with YSOs}}
We compared the cavity sizes of the transition discs around post-AGB binaries to the transition discs around YSOs. 
\citet{Varga_2018} studied 82 PPDs with VLTI/MIDI and inferred the inner disc sizes via the dust sublimation radii or with radiative transfer modelling to infer structures such as cavities and gaps.
We took their sample\footnote{https://cdsarc.cds.unistra.fr/viz-bin/cat/J/A+A/617/A83} and isolated the systems that have a dust-free cavity ($R_\mathrm{in}>2R_\mathrm{sub}$).
A comparison between the inner rim sizes of these discs and the post-AGB binary discs is shown in Fig. \ref{fig:comp_ysos}.
%We observe a clear separation in the size-luminosity diagram between the full discs and the transition discs.
The transition discs around post-AGB binaries show similar or larger physical inner rim sizes.
\modif{Due to the higher intrinsic luminosities of the post-AGB stars, $R_\mathrm{sub}$ will be larger than in YSOs (see Eq. \ref{eq:Rsub}).
The physical scale of the transition discs, which are several times $R_\mathrm{sub}$, will thus also be larger.}
In units of $R_\mathrm{sub}$, the inner disc radii are similar to the transition discs around YSOs, but show a narrower range because higher ratios imply very large physical radii (>30\,au) because the high luminosity of the post-AGB star implies large $R_\mathrm{sub}$.
%This could be because discs around post-AGB binaries more compact but ALMA continuum imaging would be needed to characterise the full extent of the dusty disc.
More complete YSO transition disc studies using ALMA (angular resolutions $\sim 0.2-0.6"$) reveal that typical dust-free cavities are $\sim 30-50$\,au \citep{vanderMarel_2023}, which is larger than $R_\mathrm{in}$ found for any of the transition discs observed with mid-IR interferometry.
The dust temperature regimes that these techniques probe are different, and larger cavities might be found for category 4 post-AGB discs.

\subsection{\modif{Origin of the dust cavities}}
\modif{The origin of the dust cavities is unclear so far. 
In YSOs, five main interpretations have been put forward to explain the large dust cavities \citep[and references therein]{vanderMarel_2023}.
These include massive planetary companions, photoevaporative clearing, dead zones, enhanced grain growth, and dynamical binary interactions.}

\modif{
For circumbinary discs around post-AGB binaries, photoevaporation and dead zones (i.e. low-ionisation regions in which the activity of magneto-rotational instability is quiescent) are unlikely to be the origin of the large inner cavities because there are no high-energy photons.
These photons are needed to cause the dust to heat and evaporate or to magnetise the disc.
}

\modif{
\citet{Birnstiel_2012} showed that the lack of near-IR excess might be caused by effective grain growth in the inner disc, depleting the $\text{micro}$metre-sized grains, but not the millimetre-sized grains.
In this scenario, inward radial drift must have occurred, which is unlikely to be efficient over the short lifetime of post-AGB discs ($\sim 10^{5}$\,yr).
However, both millimetre observations and radiative transfer modelling are needed to conclude about the possibility of enhanced grain growth as the origin of transition discs among the circumbinary disc population around post-AGB binaries.}

\modif{
The scenario in which a massive planetary companion opens the gap remains a plausible explanation and would simultaneously explain the observed cavities and the depletion pattern.
This scenario needs to be tested by taking the MATISSE phase information into account and subsequently using hydrodynamical simulations to reproduce the data.}

\modif{
The binary nature also remains a possibility because dynamical interactions induced by the binary might create the cavity.
Whether these interactions can indeed create a large enough cavity depends on the dynamical truncation radius of the binary.
In future work, the dynamical truncation radii obtained from \modiff{astrometric} orbits need to be compared with the cavity sizes obtained in this work.
The astrometric orbit is only available for AC\,Her so far \citep{Anugu_2023} and needs to be determined for all systems for a clear view.
} 

%\JK{what do you mean? there is no temperature information on the Fig. I would rather speak about physical sizes distribution and then speak about the large inner rims due to the stellar luminosity as you do a bit further.} \AC{I meant related to the size luminosity diagrams where the temperature is shown but I agree that it is not relevant here.} resulting from the lack of hot dust in the system.
%\JK{As ALMA is not probing the same dust temperatures, the existence of larger cavities around Category 4 post-AGB discs (with an excess starting even later than in the mid-IR) should be checked with ALMA.}
%Considering that the distances to the post-AGB binary systems are a factor ten larger than typical distances to near-by star forming regions, we find that the inner rim radii of post-AGB transition discs populating the higher end of the cavity sizes found for YSOs at the same angular resolution.
%Inner rim sizes of post-AGB transition discs do not populate the smaller $R_\mathrm{in}$ sizes as a result of the higher stellar luminosity.

%The observed depletion ... 
%Depletion patterns are also observed in less evolved stars \citet{Booth_Owen_2020}.

\section{Conclusion}
\label{sect:conclusion}
We presented mid-IR interferometric observations in the $L$ and $N$ bands of 11 post-AGB binary systems.
These systems are distinct in their SED because 5 systems show an IR excess starting at near-IR wavelengths, while 6 systems show a deficit at these wavelengths \modif{and the IR excess starts at mid-IR wavelengths.}
We showed that the inner dust rims of the latter 6 systems are located at 2.5 to 7.5 times the theoretical dust sublimation radius, corresponding to dust-free cavity sizes of 3-25\,au.
We therefore confirm that CT\,Ori, EP\,Lyr, AD\,Aql, RU\,Cen, and ST\,Pup, in addition to AC\,Her, are indeed transition discs, reminiscent of those that surround young stars.
In future modelling, we wish to explore the non-symmetry and potential discontinuity by fitting MATISSE phase information. 
This will allow us to test whether a planet or the binary interaction causes the transition disc nature. 
%\JK{Why? a sentence about testing the planet hypothesis is missing here.}
Moreover, we plan to use the Spectro-Polarimetric High-contrast Exoplanet REsearch instrument on the VLT and ALMA continuum observations to characterise the full extent of the dusty disc.
Furthermore, the James Webb Space Telescope will allow us to search for the gaseous species inside the inner dust rim to characterise the depletion and accretion mechanisms.

\begin{acknowledgements}
\modif{We thank the referee for their constructive comments and suggestions that substantially improved the clarity of the paper.}
A.C. and H.V.W. acknowledge support from FWO under contract G097619N. J.K. acknowledges support from FWO under the senior postdoctoral fellowship (1281121N).
This project has received funding from the European Research Council (ERC) under the European Union Horizon Europe programme (grant agreement No. 101042275, project Stellar-MADE).
D.K. and K.A. acknowledge the support of the Australian Research Council (ARC) Discovery Early Career Research Award (DECRA) grant (DE190100813). This research was supported in part by the Australian Research Council Centre of Excellence for All Sky Astrophysics in 3 Dimensions (ASTRO 3D) through project number CE170100013.
This research has benefited from the help of SUV, the VLTI user support service of the Jean-Marie Mariotti Center\footnote{http://www.jmmc.fr/suv.htm}, with special thanks to A. Matter.
This research has made use of the Jean-Marie Mariotti Center \texttt{OIFits Explorer}\footnote{Available at http://www.jmmc.fr/oifitsexplorer} and \texttt{Aspro}\footnote{Available at http://www.jmmc.fr/aspro} services.
This paper was submitted on $\pi$-day 2023.
\end{acknowledgements}

%\begingroup

\bibliographystyle{aa} % style aa.bst
\bibliography{biblio} % your references Yourfile.bib
\begin{appendix}
\section{MATISSE data reduction and calibration}
\label{sect:datareduction}
\begin{figure*}

\centering
\begin{minipage}{0.49\textwidth}
  \includegraphics[width=\textwidth,width=1.0
  \textwidth]{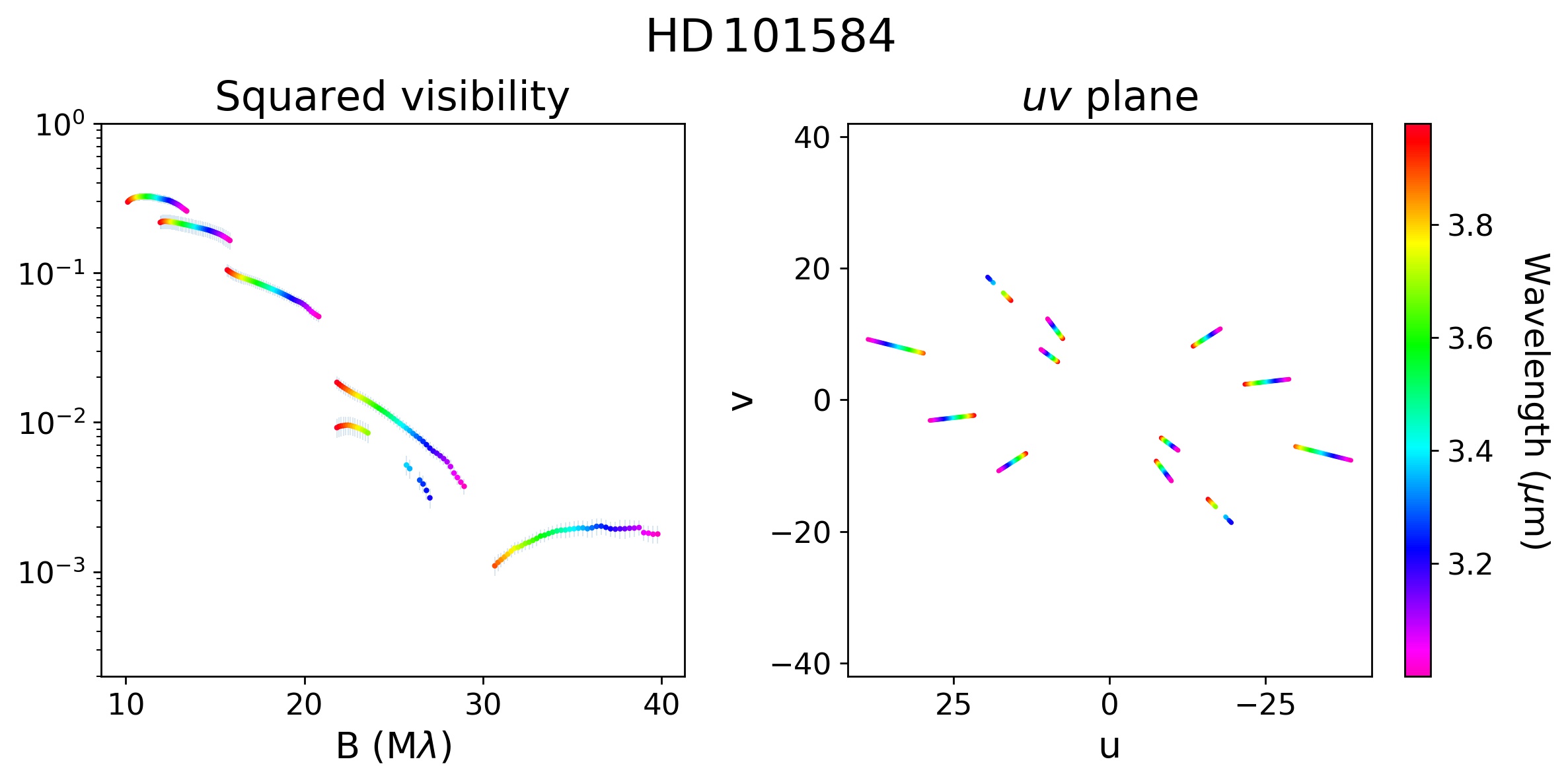}
  \end{minipage}
\begin{minipage}{0.49\textwidth}
  \includegraphics[width=\textwidth,width=1.0
  \textwidth]{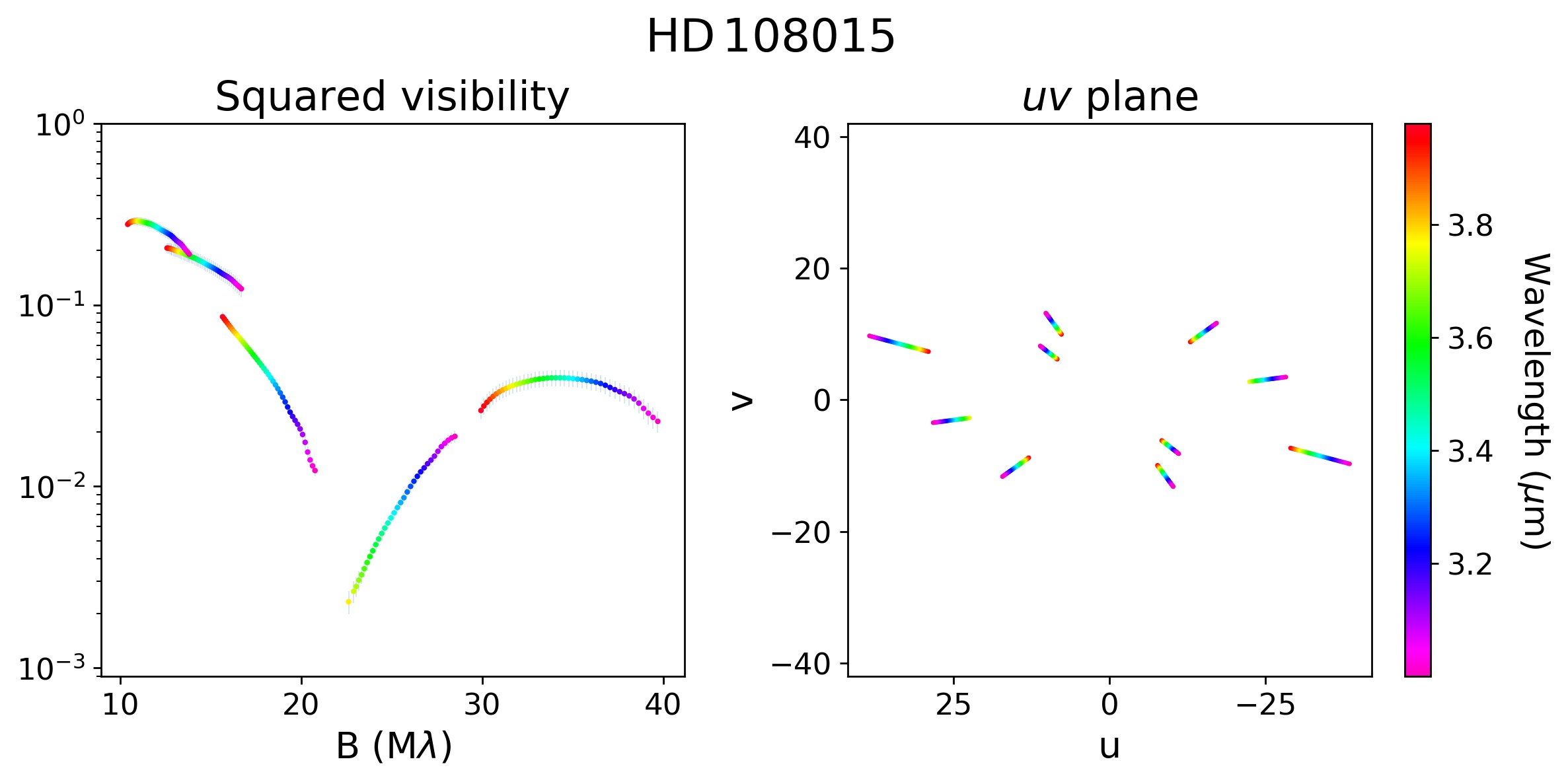}
  \end{minipage}
  \begin{minipage}{0.49\textwidth}
  \includegraphics[width=\textwidth,width=1.0
  \textwidth]{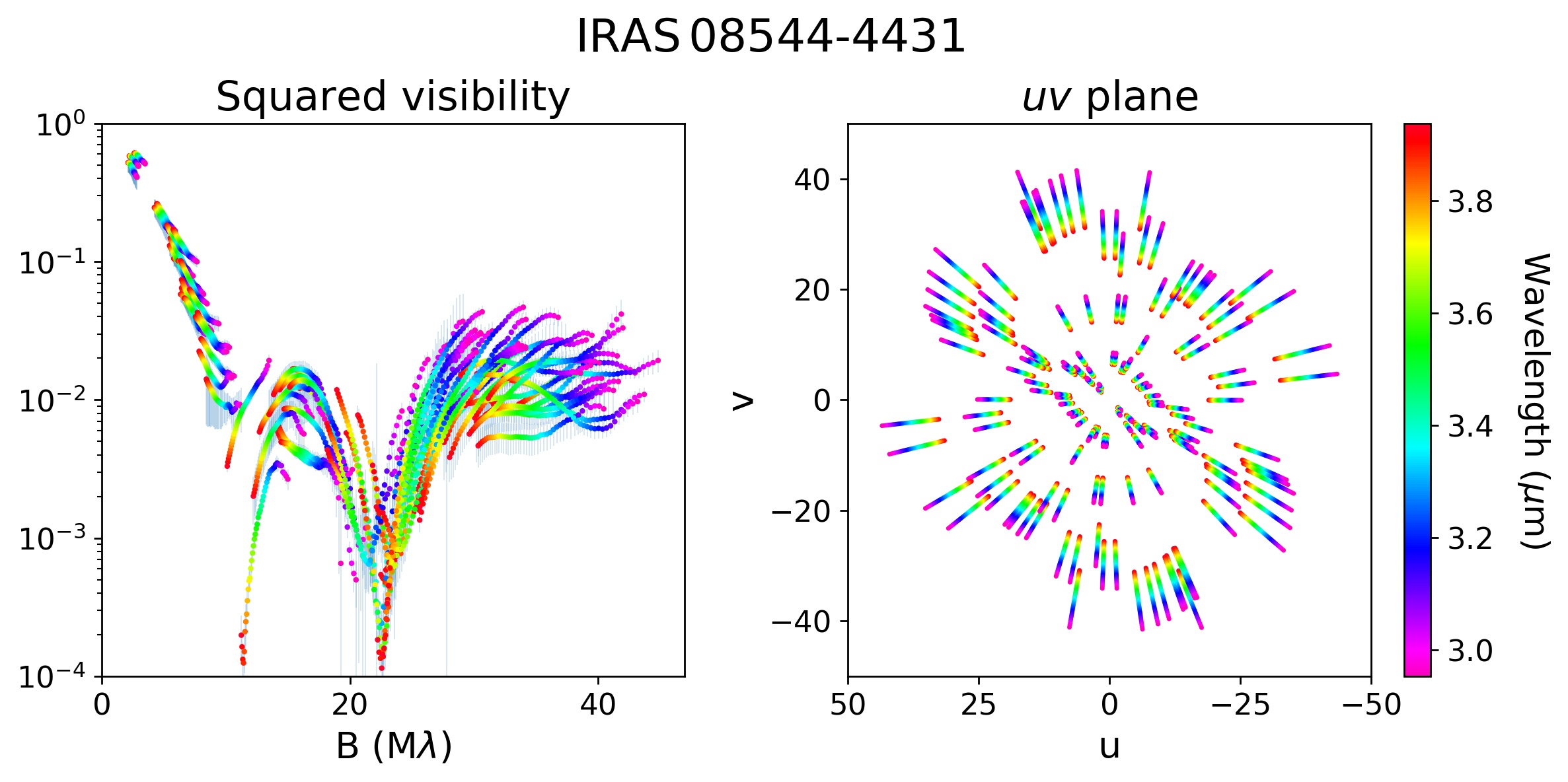}
  \end{minipage}
  \begin{minipage}{0.49\textwidth}
  \includegraphics[width=\textwidth,width=1.0
  \textwidth]{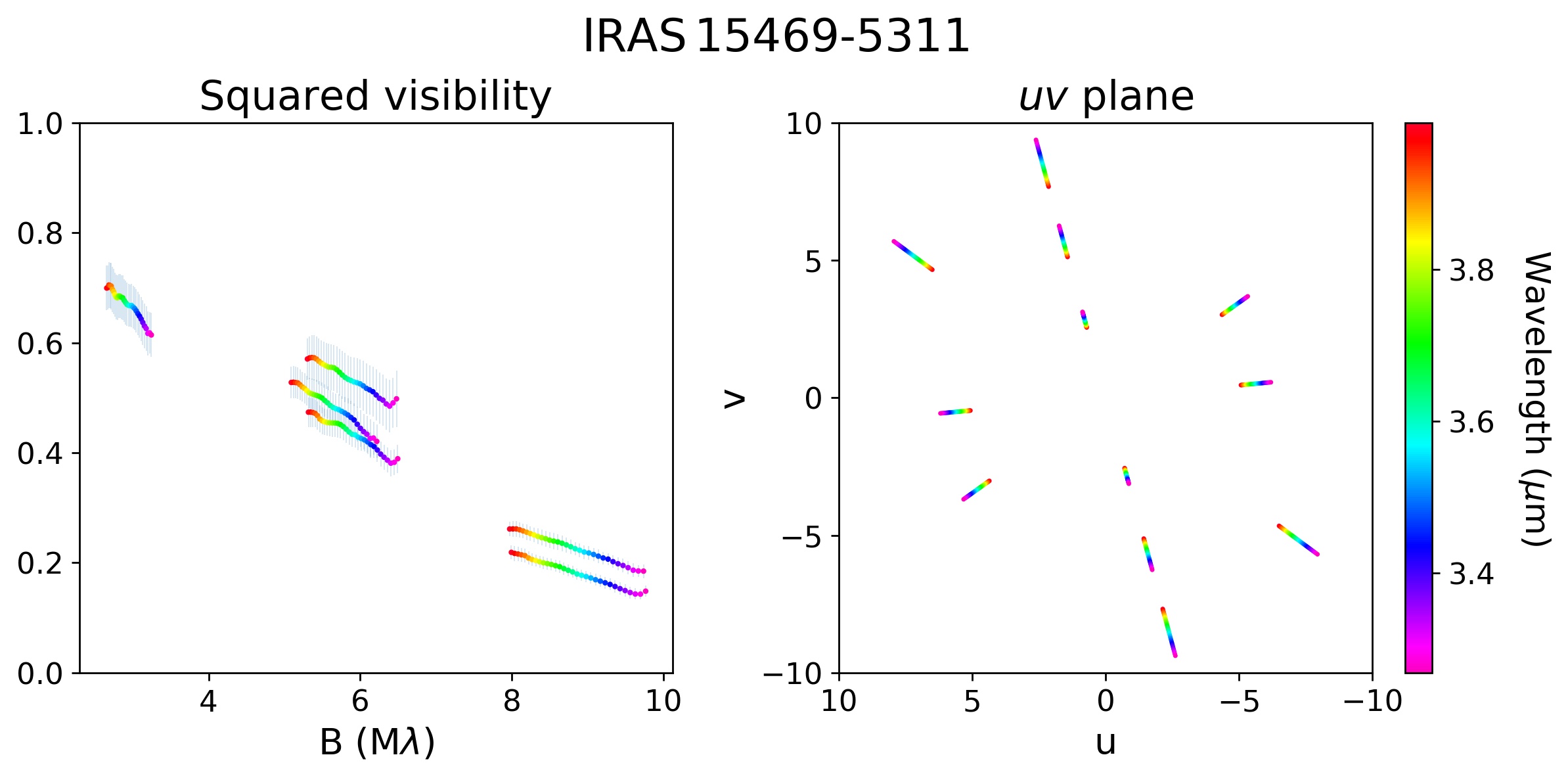}
  \end{minipage}
  \begin{minipage}{0.49\textwidth}
  \includegraphics[width=\textwidth,width=1.0
  \textwidth]{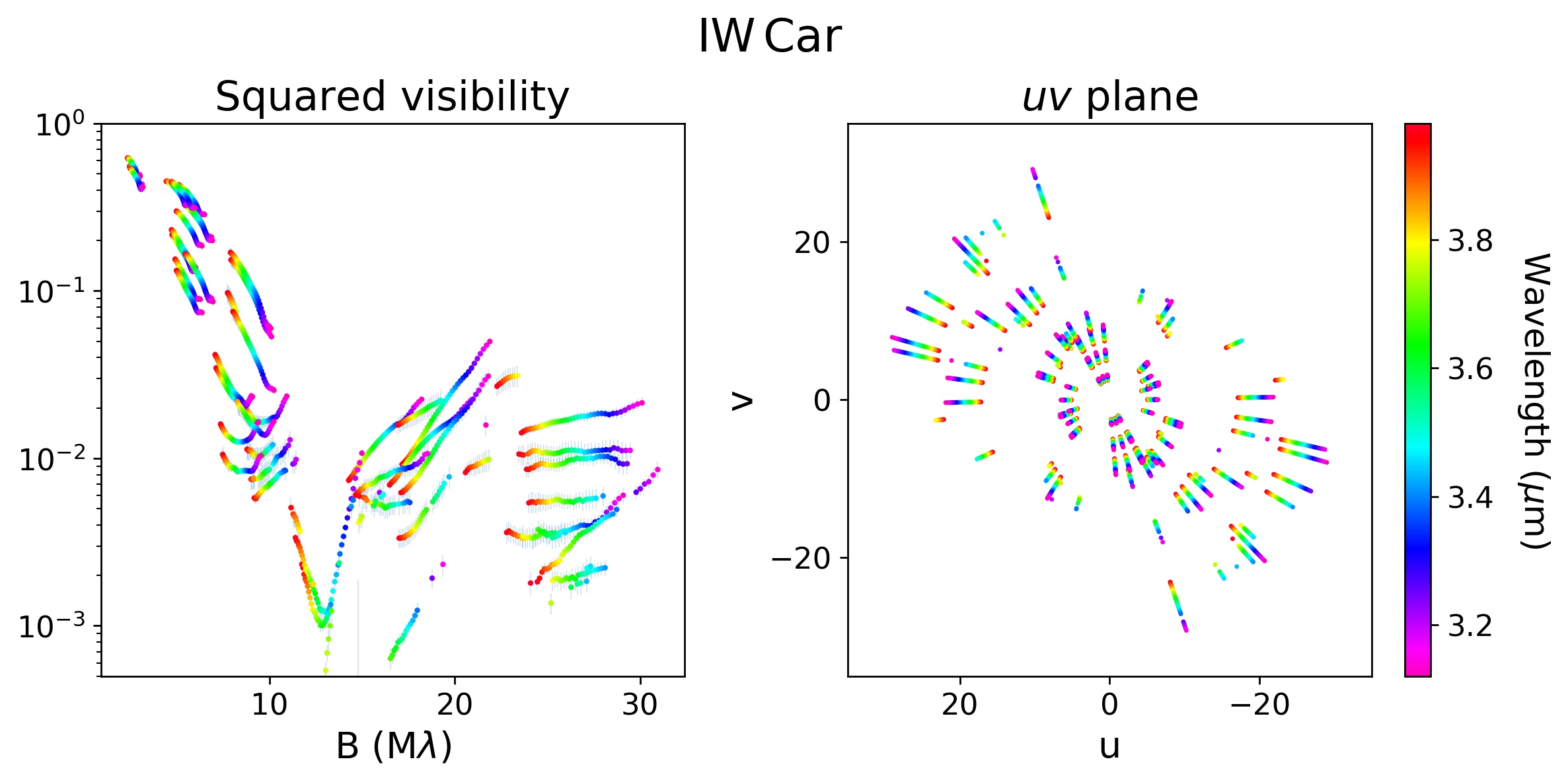}
  \end{minipage}
  \begin{minipage}{0.49\textwidth}
  \includegraphics[width=\textwidth,width=1.0
  \textwidth]{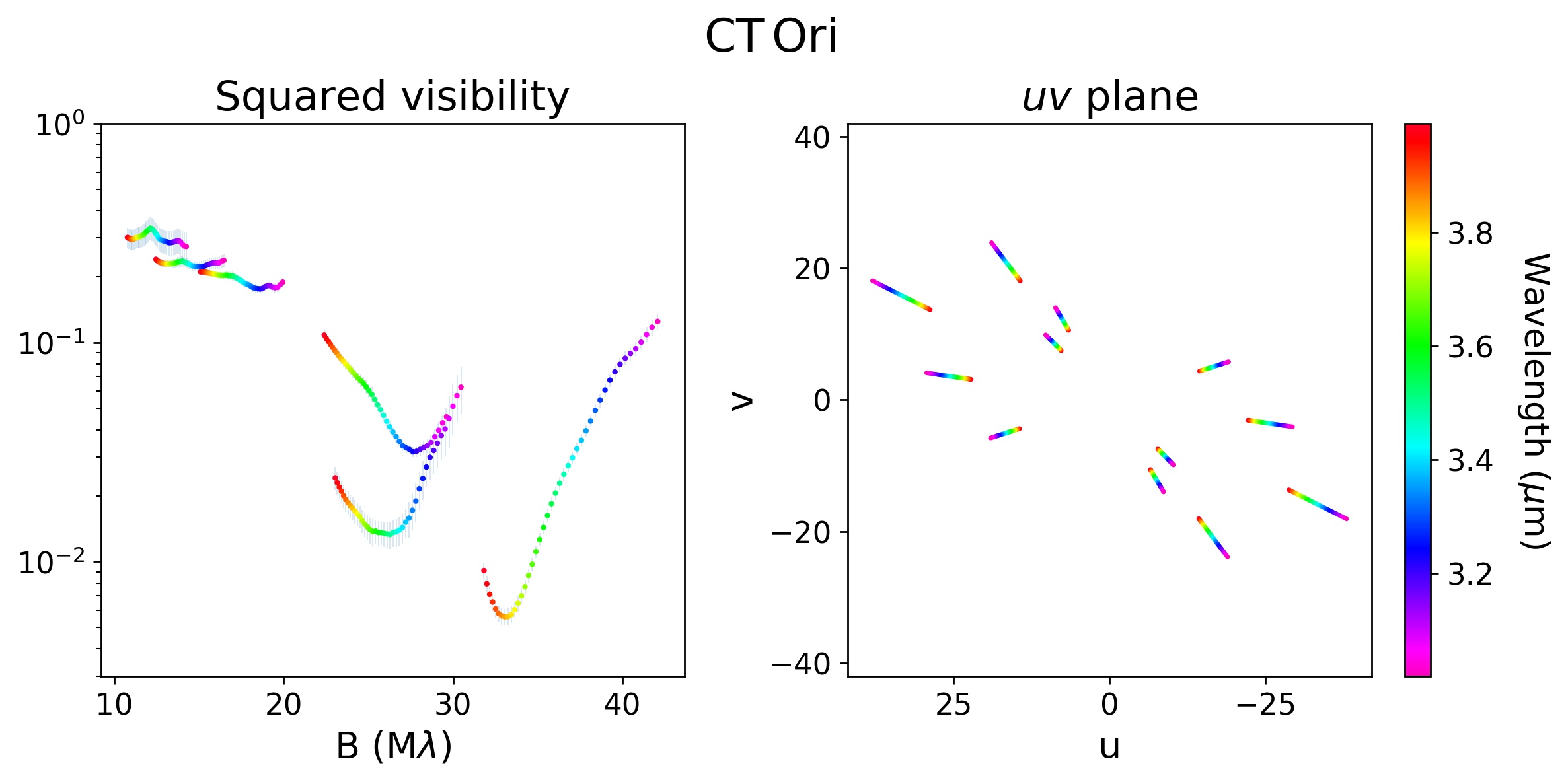}
  \end{minipage}
  \begin{minipage}{0.49\textwidth}
  \includegraphics[width=\textwidth,width=1.0
  \textwidth]{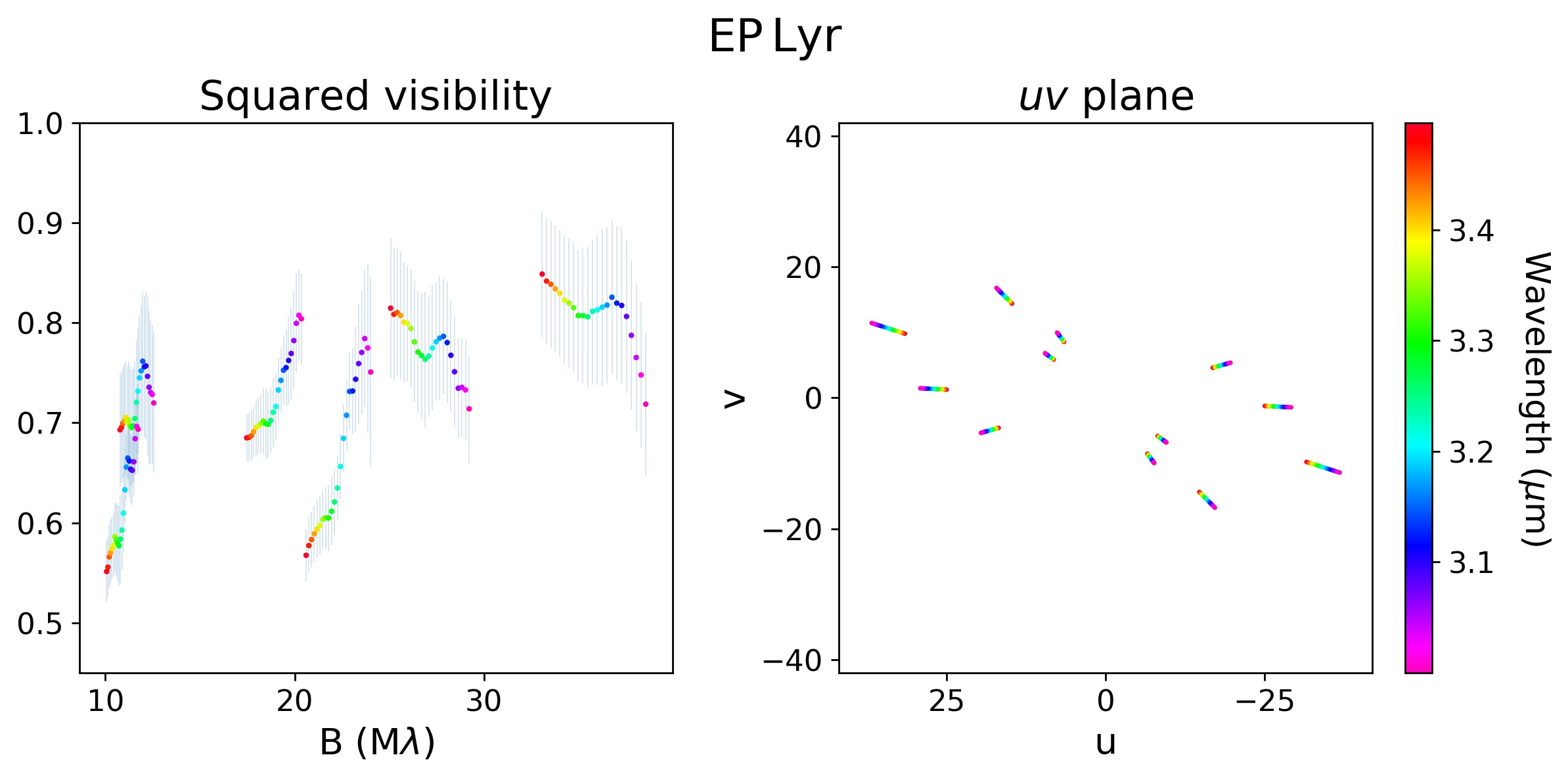}
  \end{minipage}
  \begin{minipage}{0.49\textwidth}
  \includegraphics[width=\textwidth,width=1.0
  \textwidth]{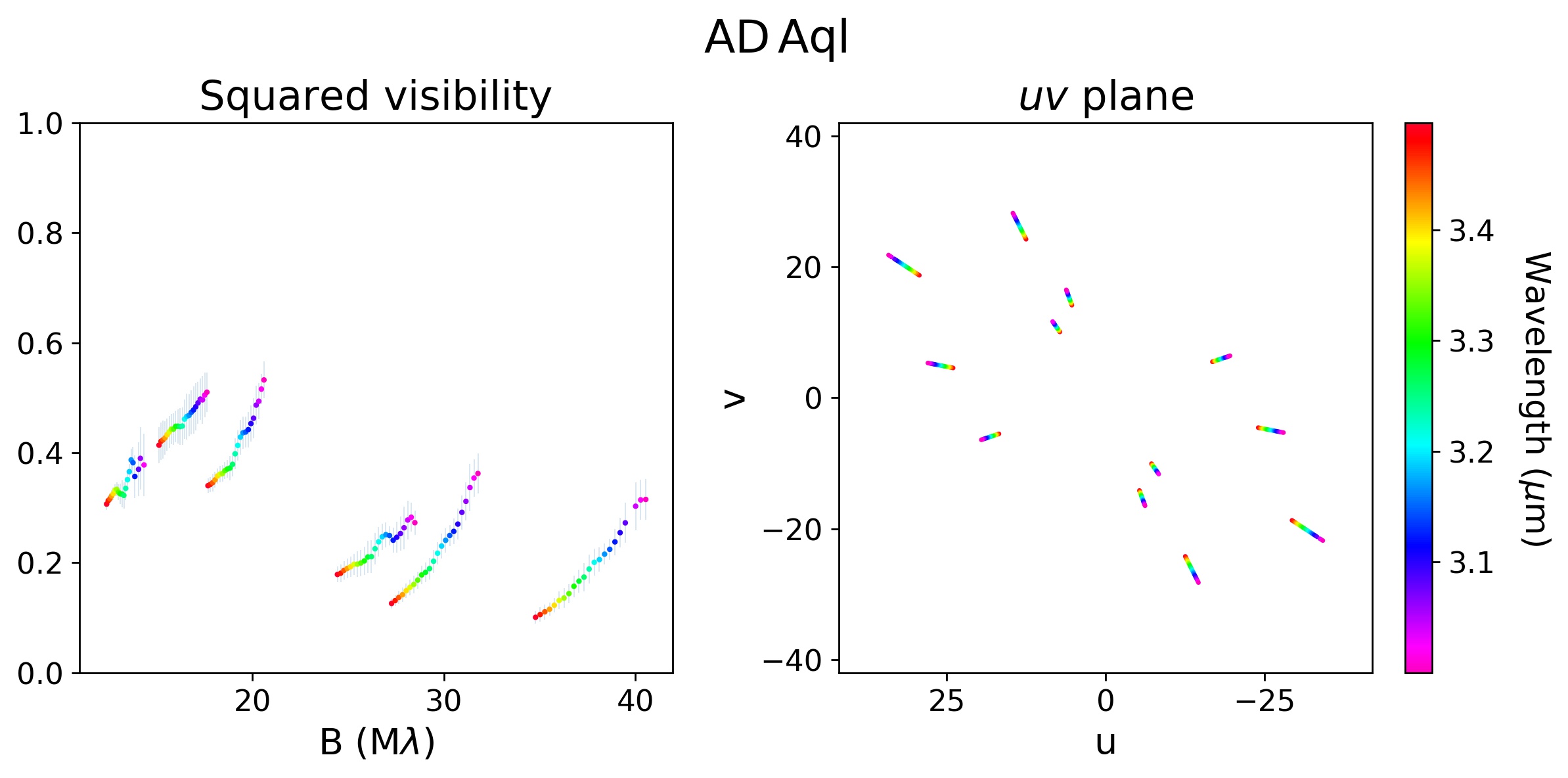}
  \end{minipage}
    \begin{minipage}{0.49\textwidth}
  \includegraphics[width=\textwidth,width=1.0
  \textwidth]{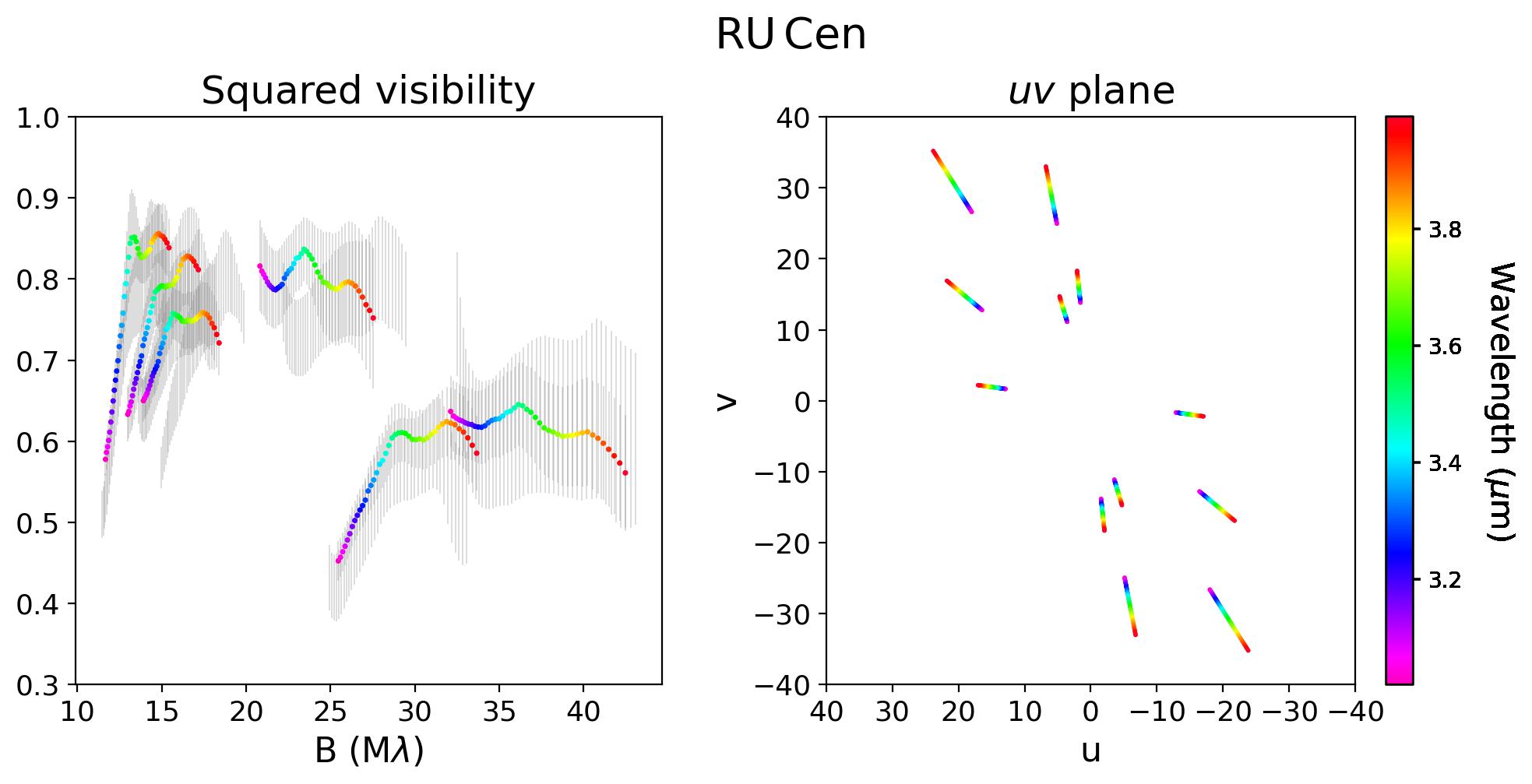}
  \end{minipage}
  \begin{minipage}{0.49\textwidth}
  \includegraphics[width=\textwidth,width=1.0
  \textwidth]{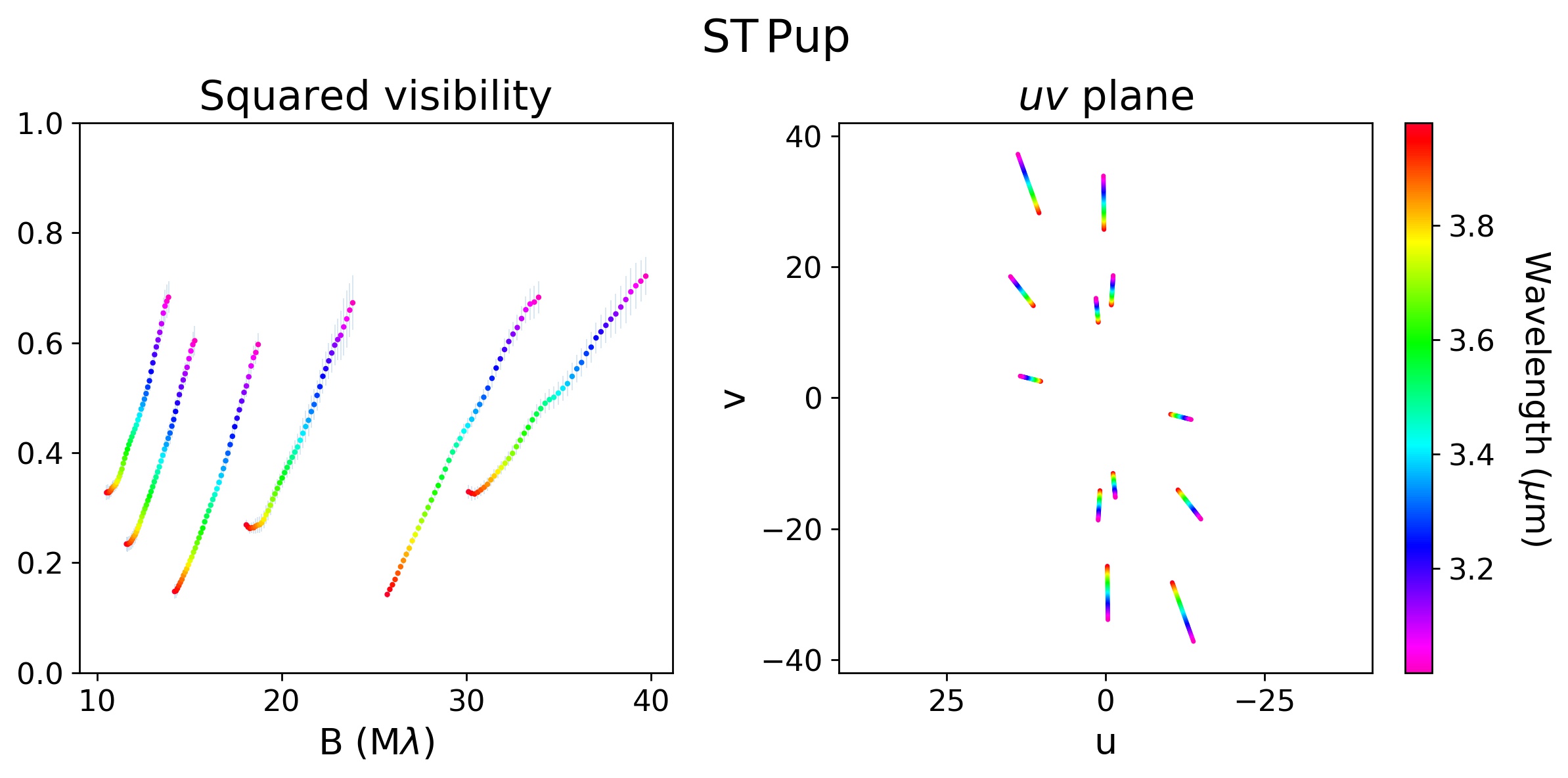}
  \end{minipage}

  \caption{Calibrated squared visibility data in the $L$ band as a function of baseline and wavelength, and the corresponding $uv$ coverages.}
  \label{fig:model_data_Lband}
\end{figure*}

\begin{figure*}

\centering
\begin{minipage}{0.49\textwidth}
  \includegraphics[width=\textwidth,width=1.0
  \textwidth]{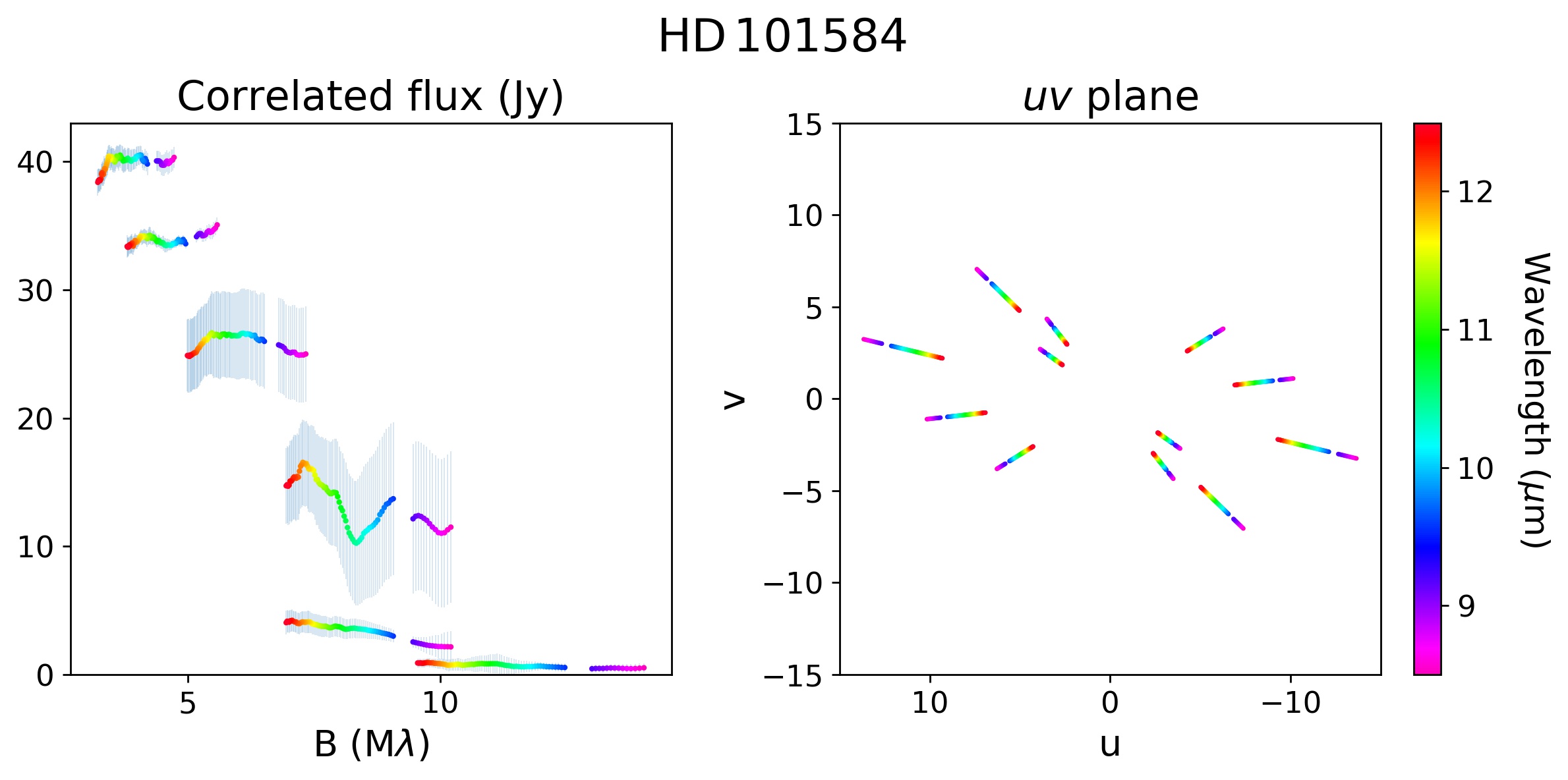}
  \end{minipage}
\begin{minipage}{0.49\textwidth}
  \includegraphics[width=\textwidth,width=1.0
  \textwidth]{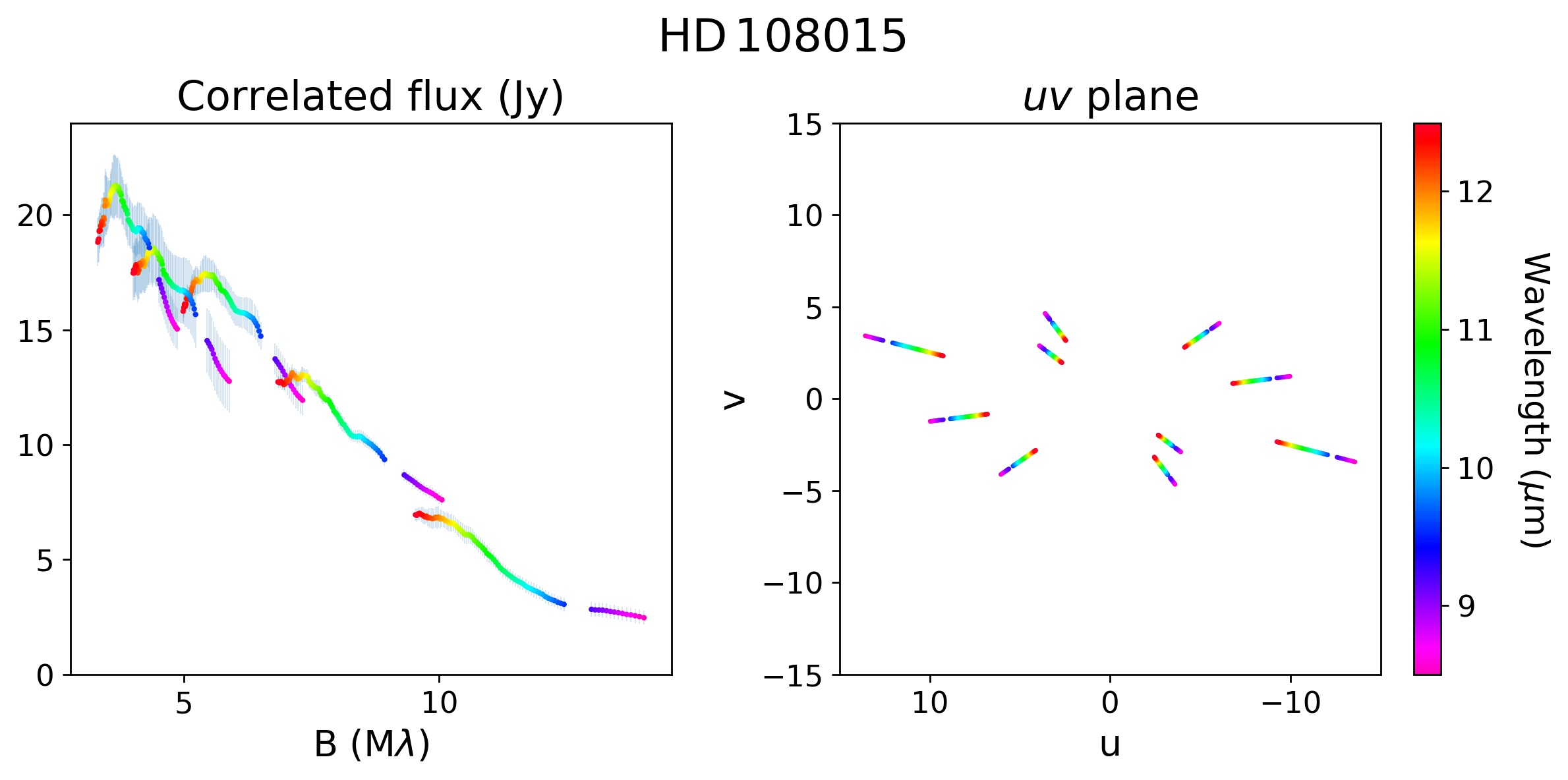}
  \end{minipage}
  \begin{minipage}{0.49\textwidth}
  \includegraphics[width=\textwidth,width=1.0
  \textwidth]{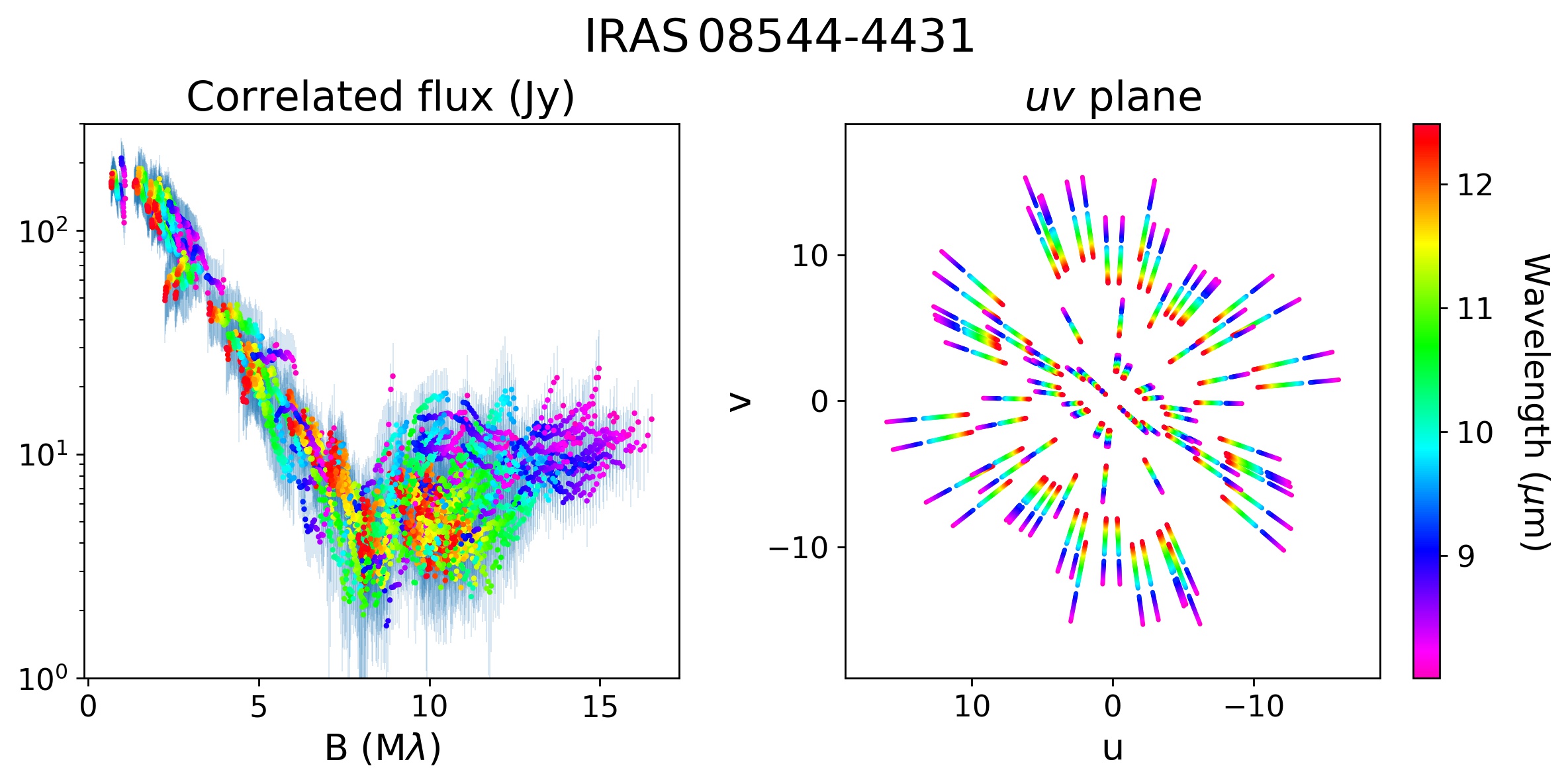}
  \end{minipage}
  \begin{minipage}{0.49\textwidth}
  \includegraphics[width=\textwidth,width=1.0
  \textwidth]{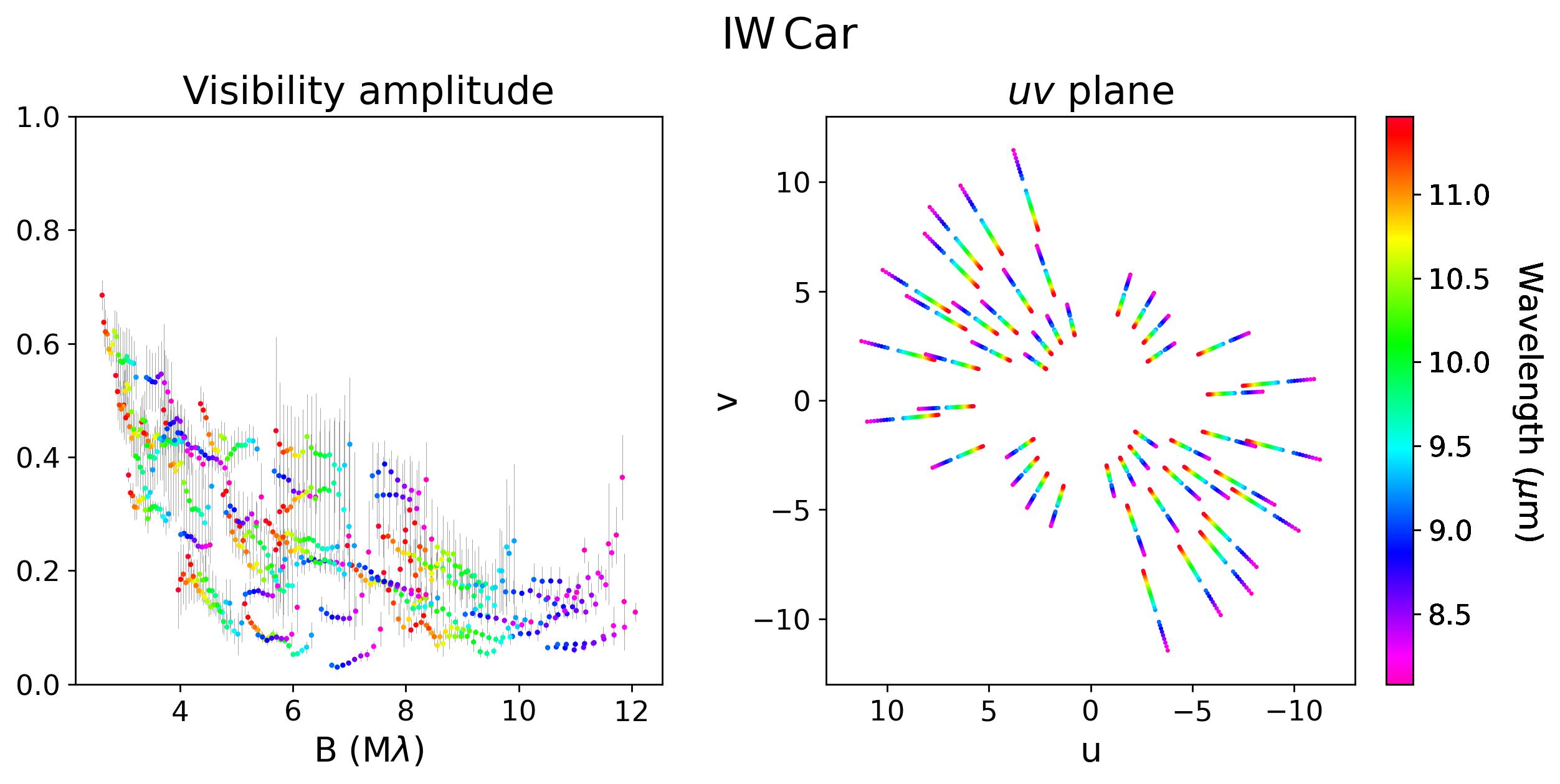}
  \end{minipage}
  \begin{minipage}{0.49\textwidth}
  \includegraphics[width=\textwidth,width=1.0
  \textwidth]{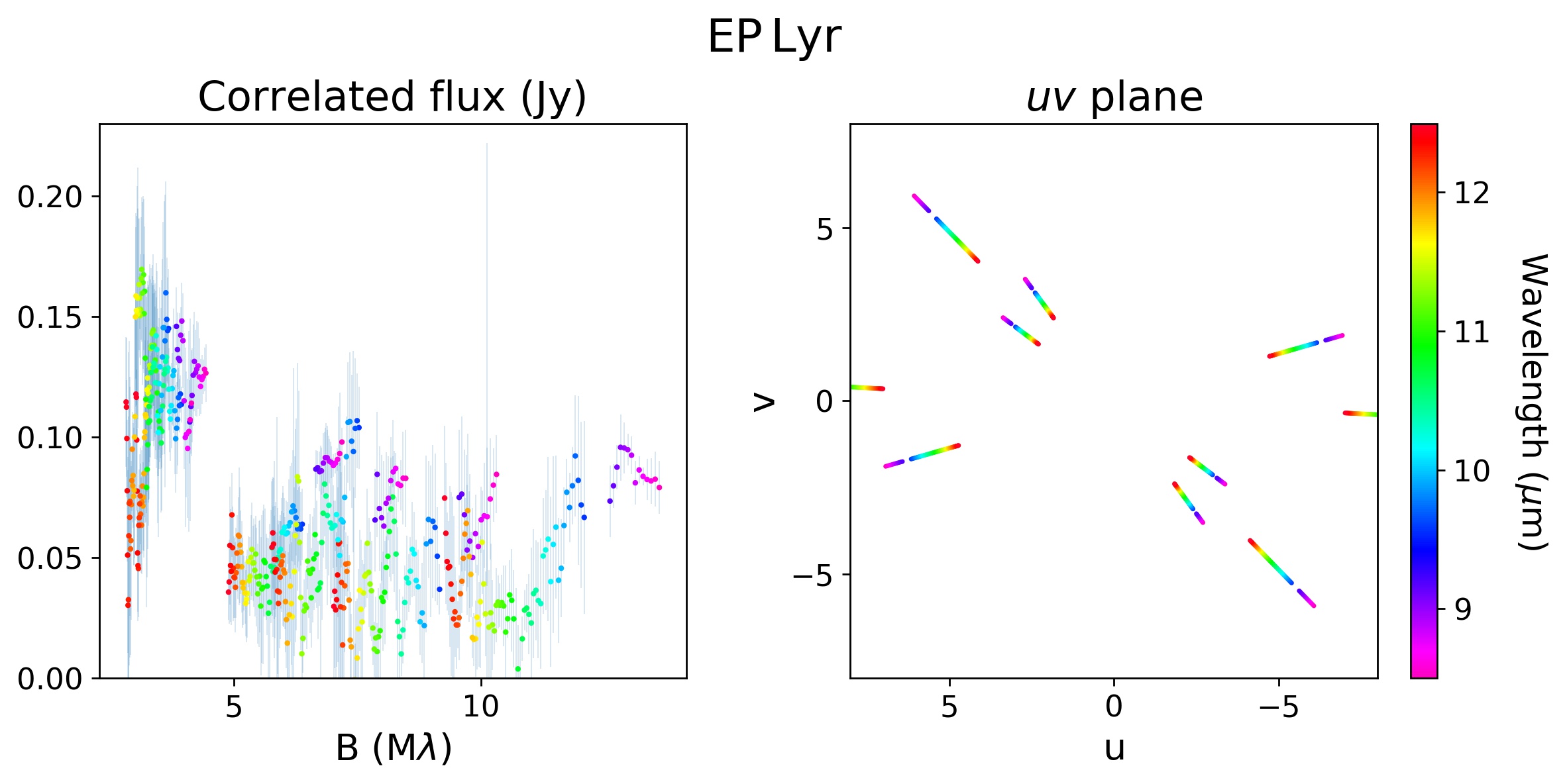}
  \end{minipage}
    \begin{minipage}{0.49\textwidth}
  \includegraphics[width=\textwidth,width=1.0
  \textwidth]{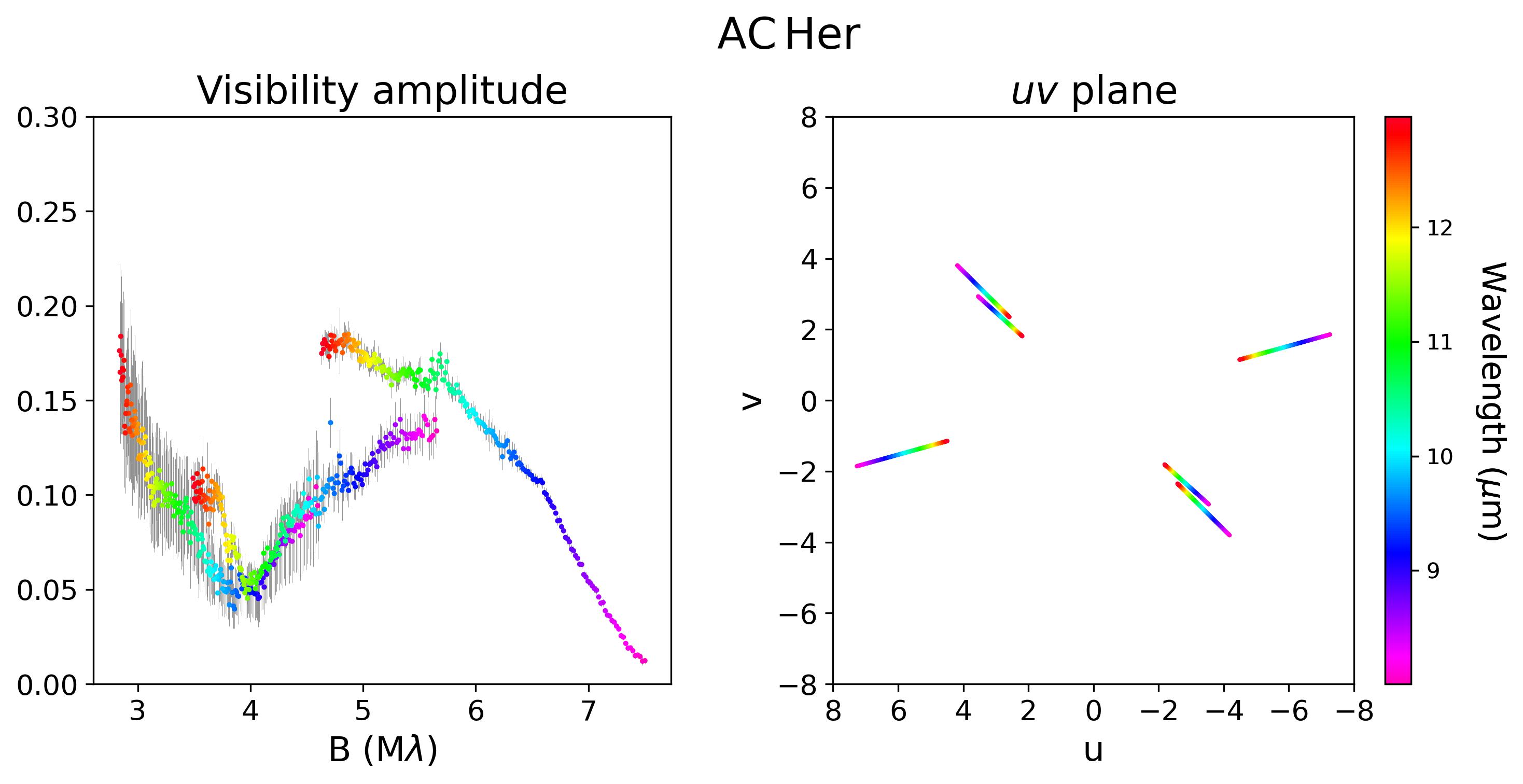}
  \end{minipage}
  \begin{minipage}{0.49\textwidth}
  \includegraphics[width=\textwidth,width=1.0
  \textwidth]{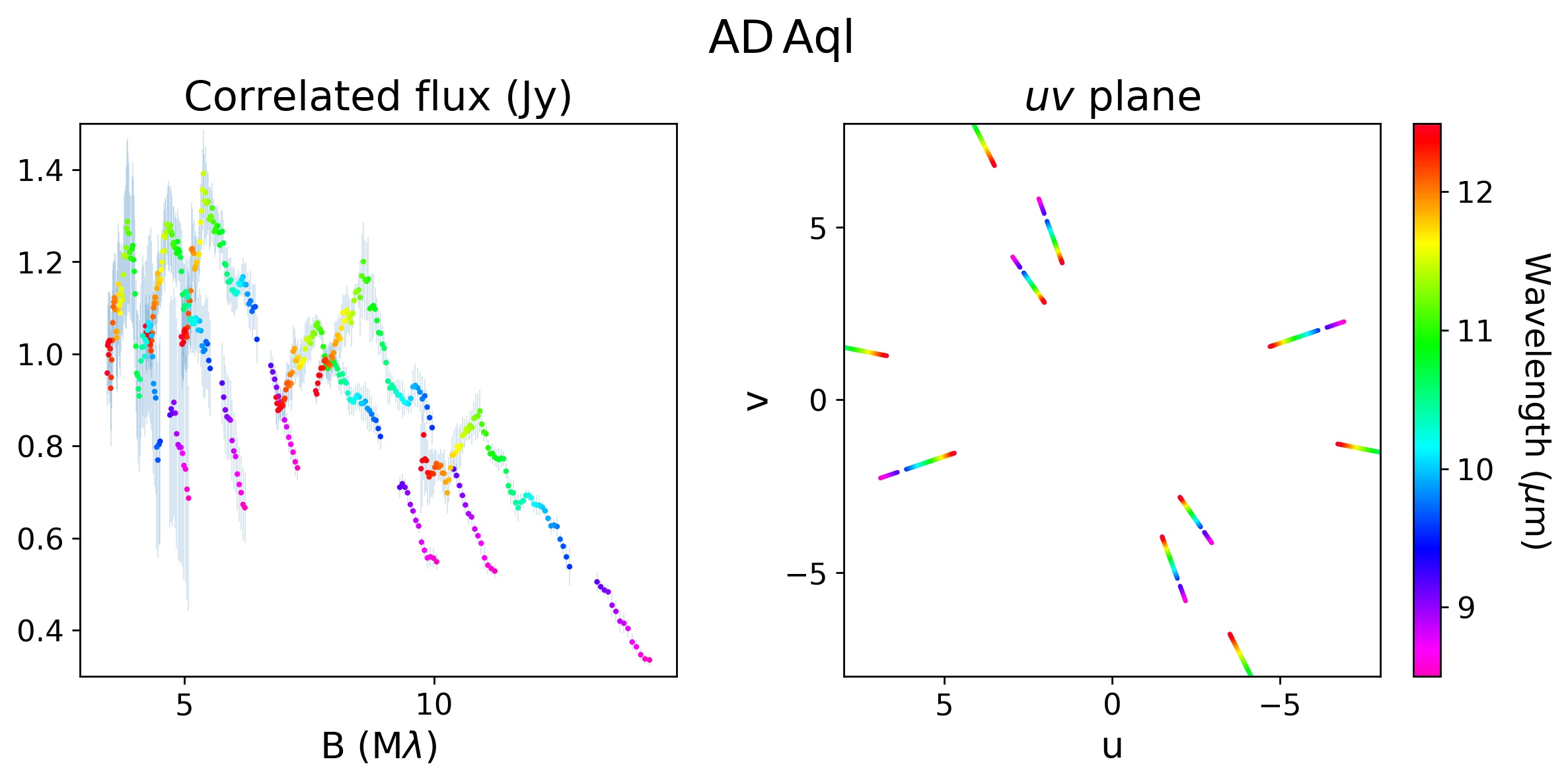}
  \end{minipage}
    \begin{minipage}{0.49\textwidth}
  \includegraphics[width=\textwidth,width=1.0
  \textwidth]{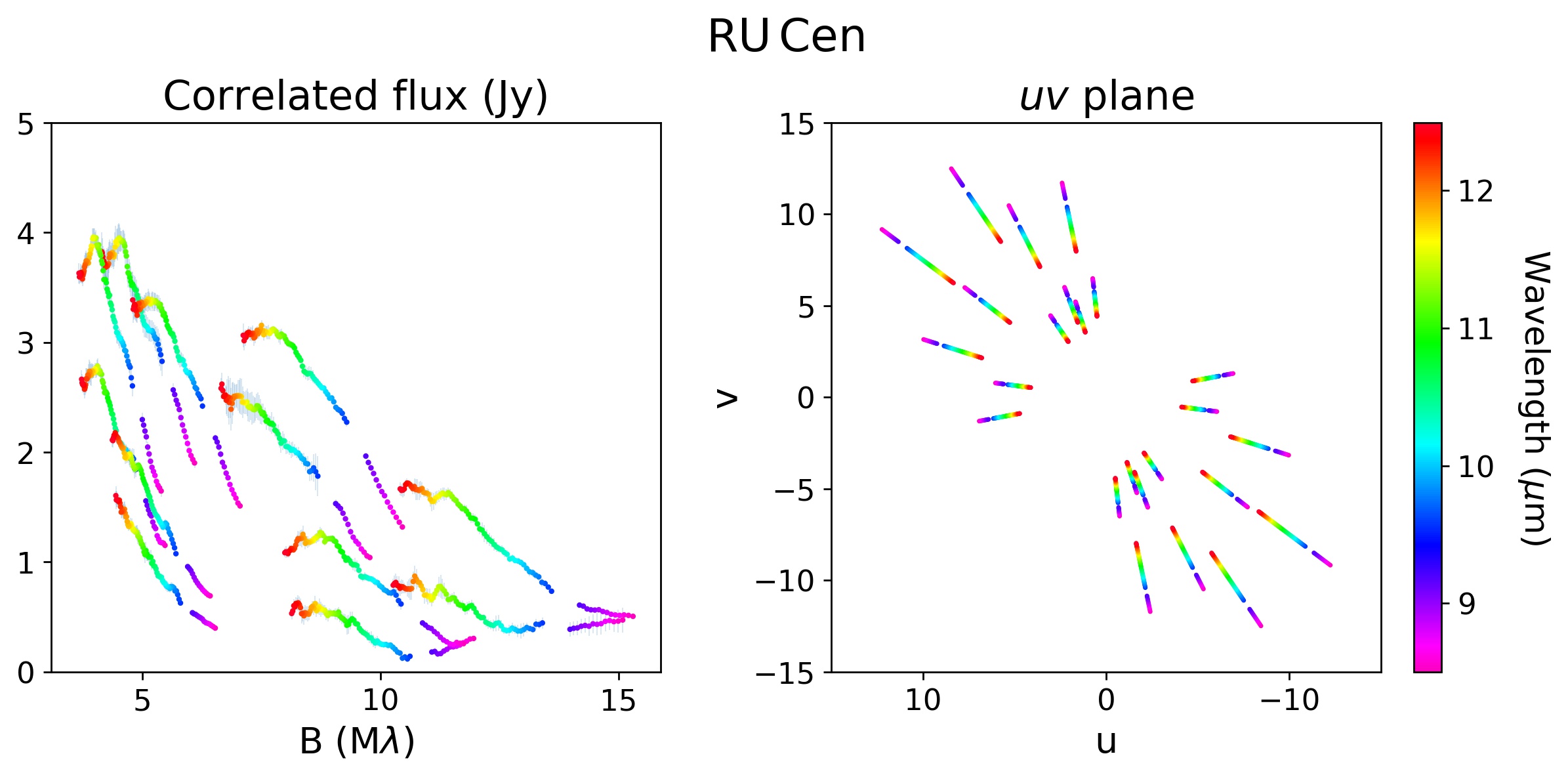}
  \end{minipage}
    \begin{minipage}{0.49\textwidth}
  \includegraphics[width=\textwidth,width=1.0
  \textwidth]{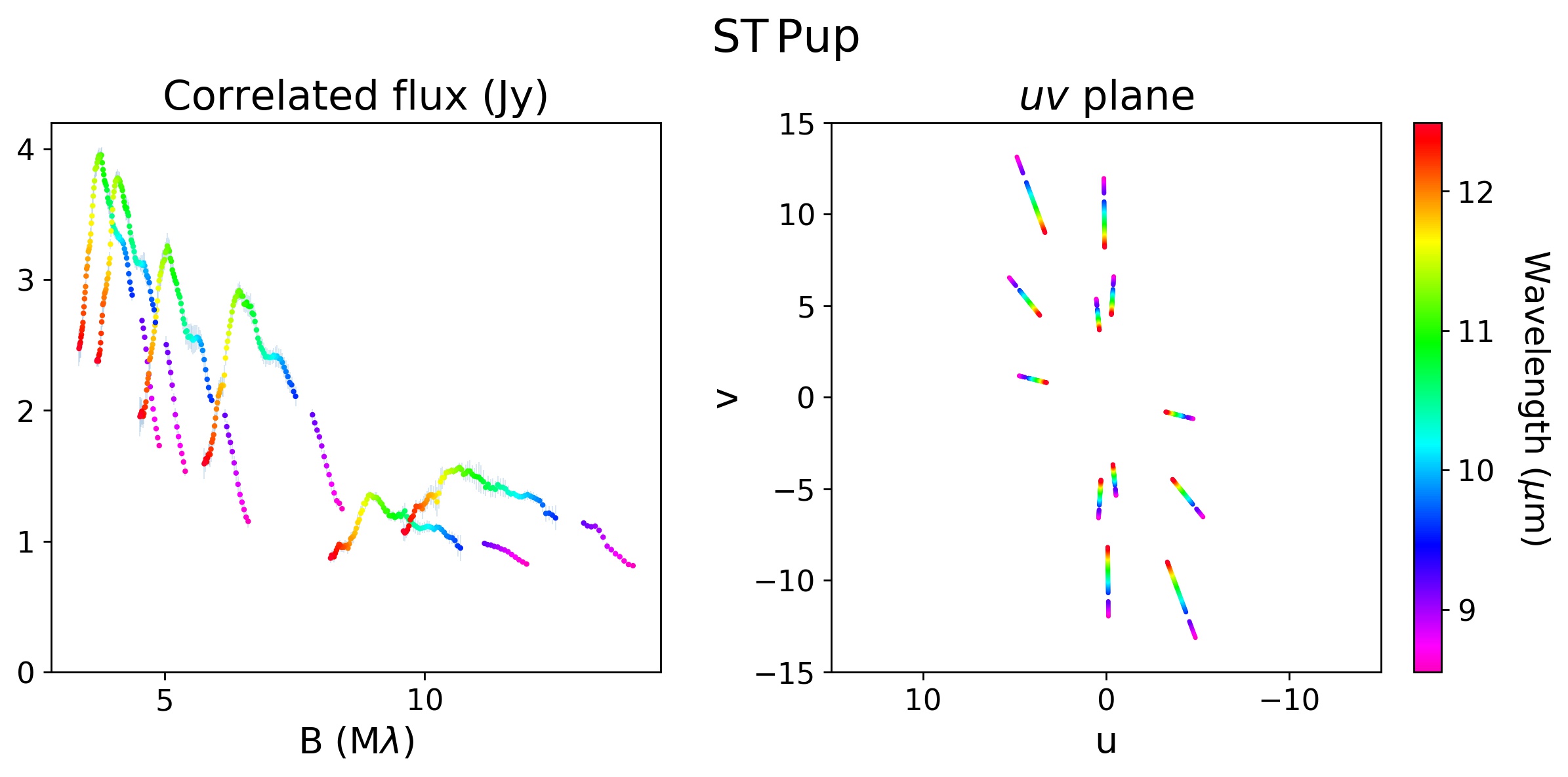}
  \end{minipage}
%     \begin{minipage}{0.24\textwidth}
%   \includegraphics[width=\textwidth,width=1.0
%   \textwidth]{Figures/RUCenring_L.jpg}
%   \end{minipage}
%   \begin{minipage}{0.24\textwidth}
%   \includegraphics[width=\textwidth,width=1.0
%   \textwidth]{Figures/STPupring_L.jpg}
%   \end{minipage}

  \caption{Calibrated visibility data in the $N$ band as a function of baseline and wavelength, and the corresponding $uv$ coverages.
  Data of HD\,101584, HD\,108015, IRAS\,08544-4431, EP\,Lyr, AD\,Aql, RU\,Cen, and ST\,Pup are reported in correlated fluxes in units of Jansky.
  Data of IW\,Car and AC\,Her are reported in visibility amplitudes.}
  \label{fig:model_data_Nband}
\end{figure*}

We outline the data reduction process and the observational and calibration problems that occurred.
A log of the observations is displayed in Table \ref{tab:MATISSE_obs}.
The calibrated $L$ - and $N$-band visibilities and the corresponding $uv$ coverages are shown in Figs. \ref{fig:model_data_Lband} and \ref{fig:model_data_Nband}, respectively.

The standard observing mode is the hybrid mode. It consists of two steps. 
First, all collected photons are sent into the interferometric and photometric channel (SiPhot mode) for the $L$ and $M$ bands and to the interferometric channel (HighSens mode) in the $N$ band.
Second, telescope chopping is done in the SiPhot mode, and photometry is taken into the $N$ band.
The second step was skipped for all observations taken with the four UTs during P109 and P110.
This means that for these observations, no chopping sequences were carried out, such that no separation of the stellar flux from the sky background could be obtained.
Coherent flux, differential phase, and closure phase measurements do not depend on chopping and could be recorded without this step.

The stability of the instrumental and atmospheric responses of observations of the science target and the calibrators were assessed from the variations in the transfer function.
Calibrators were selected from the Mid-infrared stellar Diameters and Fluxes compilation Catalogue \citep{Cruzalebes_2019} and were checked for being flagged for IR excesses or bad object types. 
\modif{The intended $N$-band calibrators for IRAS\,15469 and the small-baseline data of IW\,Car are bad calibrators as they show both an IR excess and an IR extent.}
The $L$-band calibrator could not be used to calibrate the data either because its $N$-band flux is too low.
Therefore, the $N$-band data of IRAS\,15469 could not be calibrated and were discarded.
\modif{Similarly, for IW\,Car, the small-baseline data could not be calibrated, and only the medium baseline $N$-band data were used.}

In the standard MATISSE pipeline, a spectral channel is represented by five and seven spectral pixels on the detector in the $L$ and $N$ bands, respectively.
The data were thus binned accordingly.
The data of all targets except for EP\,Lyr were reduced with these settings.
EP\,Lyr is faint, and its mid-IR brightness of 0.25\,Jy in $N$ is at the limits of the MATISSE instrument of 0.1\,Jy correlated flux that could be guaranteed in P109.
%\JK{put some numbers please}.
The resulting S/N was not enough with a regular spectral binning in the $N$ band.
To increase the S/N, the data were reduced using a spectral binning of 21 pixels.
The $N$-band data of IW\,Car were spectrally binned \modif{by a factor of three} after data reduction to increase the S/N.

During the observations of HD\,108015, the baseline UT1-UT3 had technical issues that significantly affected the $N$-band correlated flux.
Consequently, baseline UT1-UT3 was disregarded in the modelling for this target.

For the $L$ band, the consequence of skipping the second step in the standard hybrid mode is that it lacks a correction of the thermal background fluctuations because no chopping is done.
This mainly affected the data of EP\,Lyr and AD\,Aql.
For these objects, photometry taken in $L$ in the first step was not reliable at $\lambda>3.5$\,$\mu$m because they were affected by temporal fluctuations in the atmospheric conditions.
Therefore, data at $\lambda>3.5\,\mu$m were discarded for these two targets.
Other (changes in) atmospheric conditions likely affected the observations of CT\,Ori. 
The $N$-band data of CT\,Ori were discarded because they were found to be very noisy.

\begin{table*}
\caption{Log of MATISSE observations}
\label{tab:MATISSE_obs}
\centering
\begin{threeparttable}
\begin{tabular}{lcccccc}
\hline \hline
Object & Date & Progam ID & MJD & Configuration & Calibrator(s) L & Calibrator(s) N\\
\hline
HD\,101584 & 2022-05-23& 109.23KF&  59722.06&  UT1UT2UT3UT4& V918\,Cen & V918\,Cen\\
HD\,108015& 2022-05-23& 109.23KF& 59722.10& UT1UT2UT3UT4 & W\,Cen&W\,Cen \\
IRAS\,15469 &  2021-05-21 & 105.20QN & 59355.12 &  A0B2D0C1 & HD\,140354 & - \\
IW\,Car (A)&2020-12-30  &106.21NK&59213.24 & A0B2D0C1 & HD\,77324 + CD-553254 & -\\
\modif{IW\,Car (B)}&2020-12-30  &106.21NK&59213.30 & A0B2D0C1 & HD\,77324 + CD-553254 & -\\
\modif{IW\,Car (C)}& 2021-01-02 & 106.21NK & 59216.29 & A0B2D0C1 & HD\,77324 + CD-553254 & -\\
\modif{IW\,Car (D)} & 2021-01-03 & 106.21NK & 59217.18 & A0B2D0C1 & HD\,77324 + CD-553254 & -\\
\modif{IW\,Car (E)} & 2021-02-23 & 106.21NK & 59268.16 & K0G2D0J3 & HR\,3914  & N\,Vel\\
\modif{IW\,Car (F)} & 2021-02-26 & 106.21NK & 59271.05 &  K0G2D0J3& HR\,3914  & N\,Vel \\ 
\modif{IW\,Car (G)} & 2021-03-10 & 106.21NK & 59283.10  &  K0G2D0J3& HR\,3914 & N\,Vel\\ 
\modif{IW\,Car (H)} & 2021-03-10 & 106.21NK & 59283.18 &  K0G2D0J3& HR\,3914 & N\,Vel\\ 
\modif{IW\,Car (I)} & 2021-03-20 & 106.21NK & 59293.05 &  K0G2D0J3& HR\,3914 & N\,Vel\\ 
\modif{IW\,Car (J)} & 2021-03-20 & 106.21NK & 59293.10 &  K0G2D0J3& HR\,3914 & N\,Vel\\ 

CT\,Ori & 2023-01-07&110.24AC & 59951.17& UT1UT2UT3UT4 & 17\,Mon & 17\,Mon \\
EP\,Lyr & 2022-05-23& 109.23KF&59722.32 & UT1UT2UT3UT4 & HD\,180450& HD\,180450 \\
AD\,Aql & 2022-05-23& 109.23KF& 59722.27& UT1UT2UT3UT4 & i Aql&alf Sct \\
RU\,Cen (A) & 2022-05-22& 109.23KF& 59722.00& UT1UT2UT3UT4 & B\,Cen& HD\,112213 \\
RU\,Cen (B) &2023-01-08 & 110.24AC&59952.30 & UT1UT2UT3UT4 &B\,Cen& HD\,112213  \\
ST\,Pup &2022-11-15 & 110.24AC& 59864.27& UT1UT2UT3UT4 & HD\,50235 & HD 47536\\
\hline
\end{tabular}
\end{threeparttable}
\end{table*}

\section{System properties}
Table \ref{table:stellarprops} shows the properties of the 11 systems in the sample.
To summarise, category 1 targets are surrounded by a full disc and show larger [Fe/H] and lower [Zn/Ti] than category 2 and 3 targets, which are surrounded by transition discs.
\begin{table*}
\caption{Stellar properties}
\label{table:stellarprops}
\centering
\begin{threeparttable}
\begin{tabular}{lcccccc}
\hline \hline
Object & Cat. & $T_\mathrm{eff}$ (K)& [Fe/H] & [Zn/Ti] & RVb & $T_\mathrm{turn-off}$ (K) \\
\hline
HD\,101584 & Cat. 1 & 7000 & 0.0 & - & no& -\\
HD\,108015 & Cat. 1 & 7000 & -0.1 & 0.10 & no & -\\
IRAS\,08544 & Cat. 1 & 7250 & -0.5& 0.90 & no & 1200\\
IRAS\,15469 & Cat. 1 & 7500 & 0.0 & 1.80 & no & 1300\\
IW\,Car & Cat. 1 & 6700 & -1.0 & 2.1 & yes& 1100\\
CT\,Ori & Cat. 3 & 5500 & -2.0 & 1.90 & no& 1200\\
EP\,Lyr & Cat. 3& 7000& -1.5 & 1.40 & yes& 800\\
AC\,Her & Cat. 2 & 5500 & -1.7 & 1.00 & no & 1200\\
AD\,Aql & Cat. 2& 6300 & -2.0 & 2.50 & no& 1000\\
RU\,Cen & Cat. 2 & 6000 & -2.0 & 1.00 & no & 800\\
ST\,Pup & Cat. 2 & 5750 & -1.5 & 2.10 & no& 800\\
\hline
\end{tabular}
\end{threeparttable}
\tablefoot{Values taken from the catalogue of \citet{Kluska_2022}.
}
\end{table*}

\section{Best-fitting models}
\begin{figure*}

\centering
\begin{minipage}{0.24\textwidth}
  \includegraphics[width=\textwidth,width=1.0
  \textwidth]{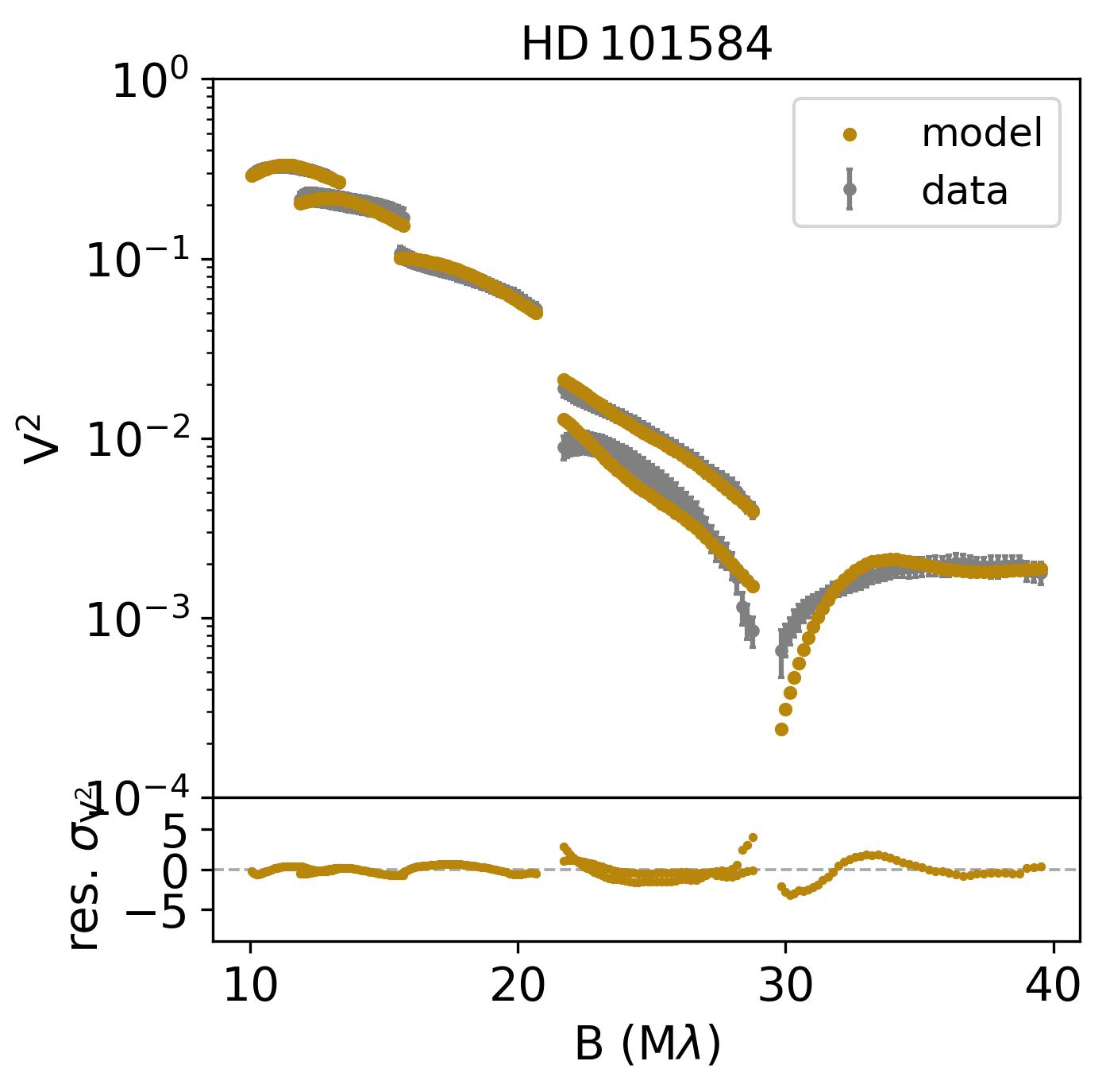}
  \end{minipage}
\begin{minipage}{0.24\textwidth}
  \includegraphics[width=\textwidth,width=1.0
  \textwidth]{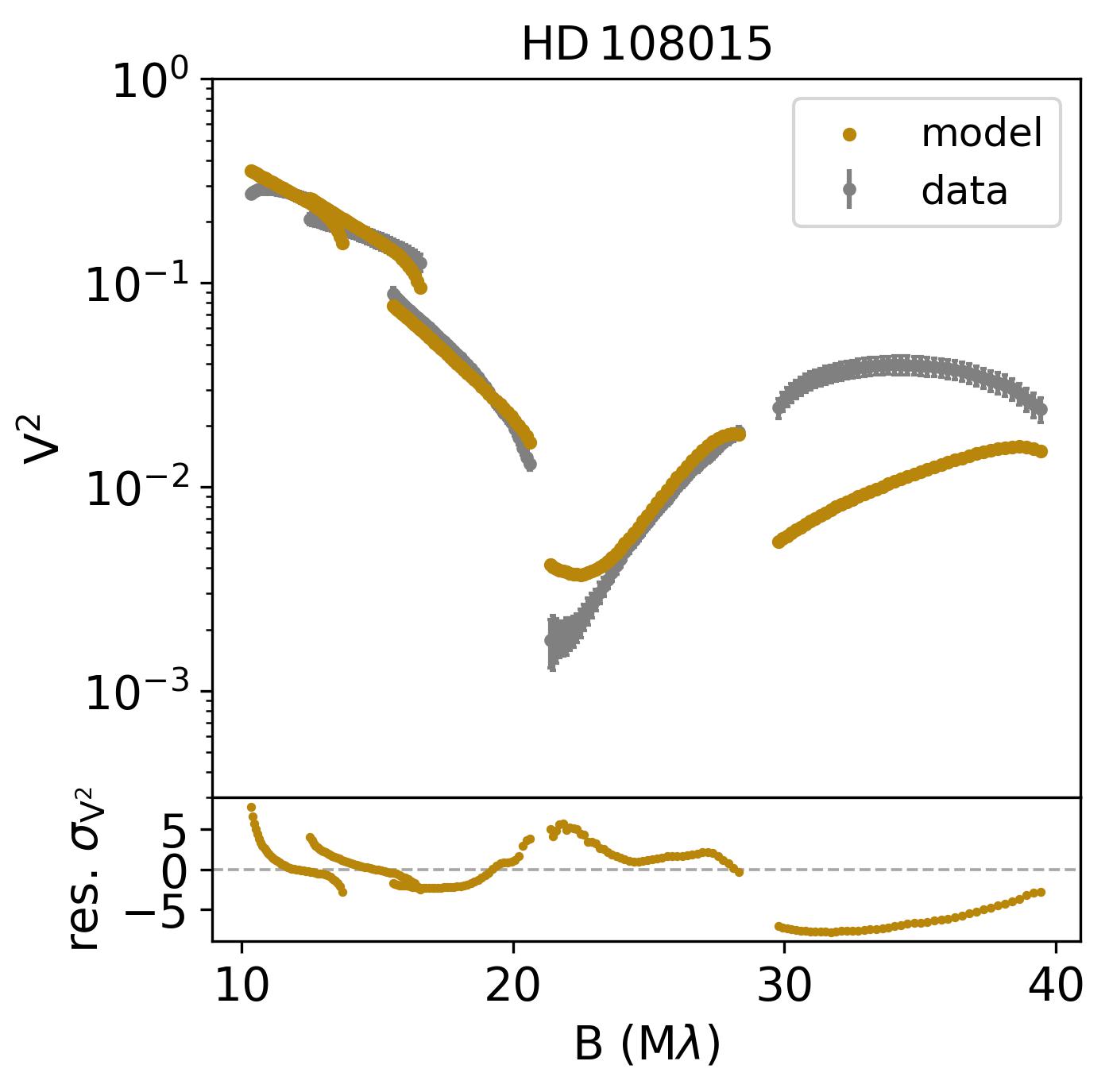}
  \end{minipage}
  \begin{minipage}{0.24\textwidth}
  \includegraphics[width=\textwidth,width=1.0
  \textwidth]{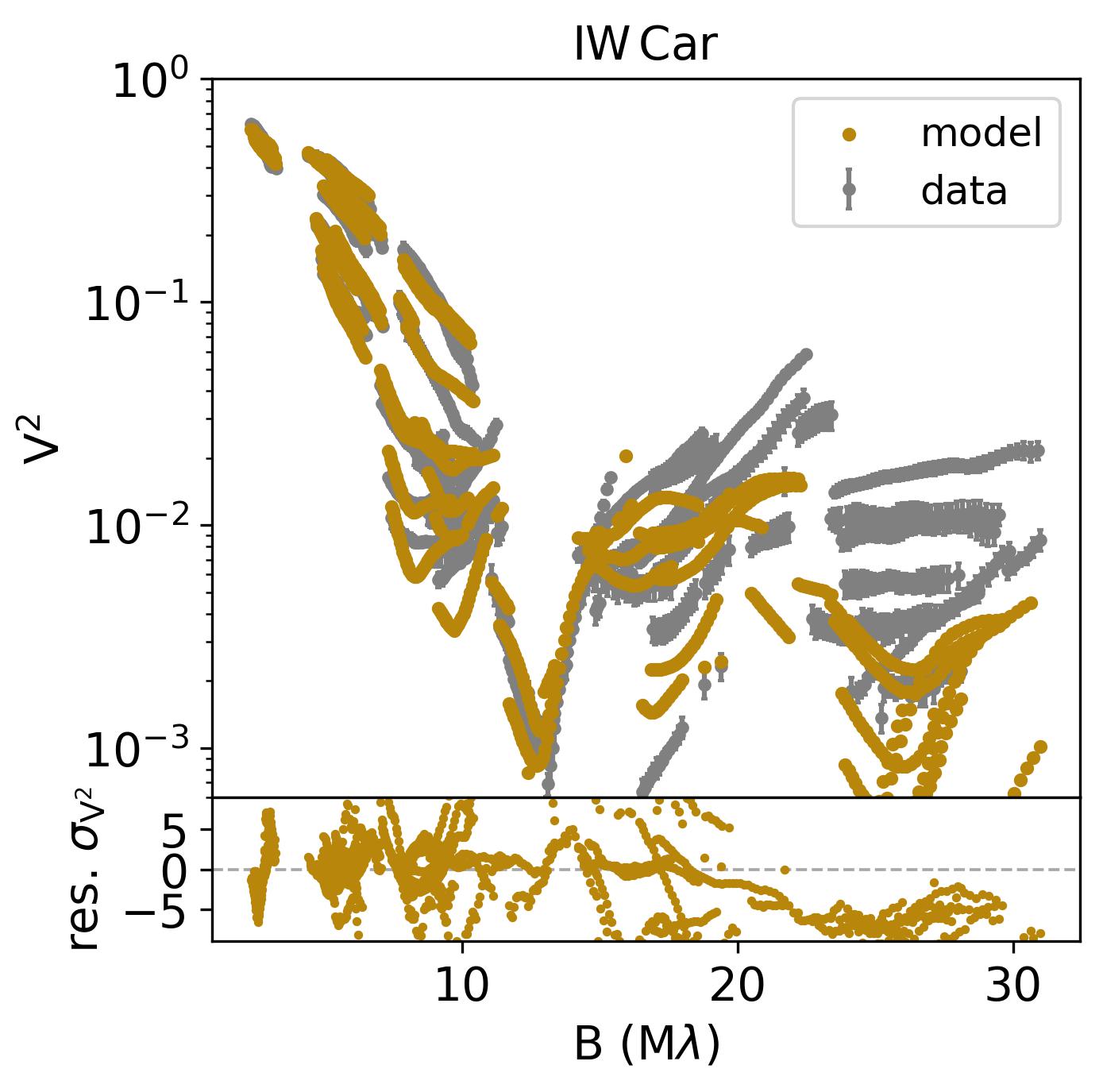}
  \end{minipage}
\begin{minipage}{0.24\textwidth}
  \includegraphics[width=\textwidth,width=1.0
  \textwidth]{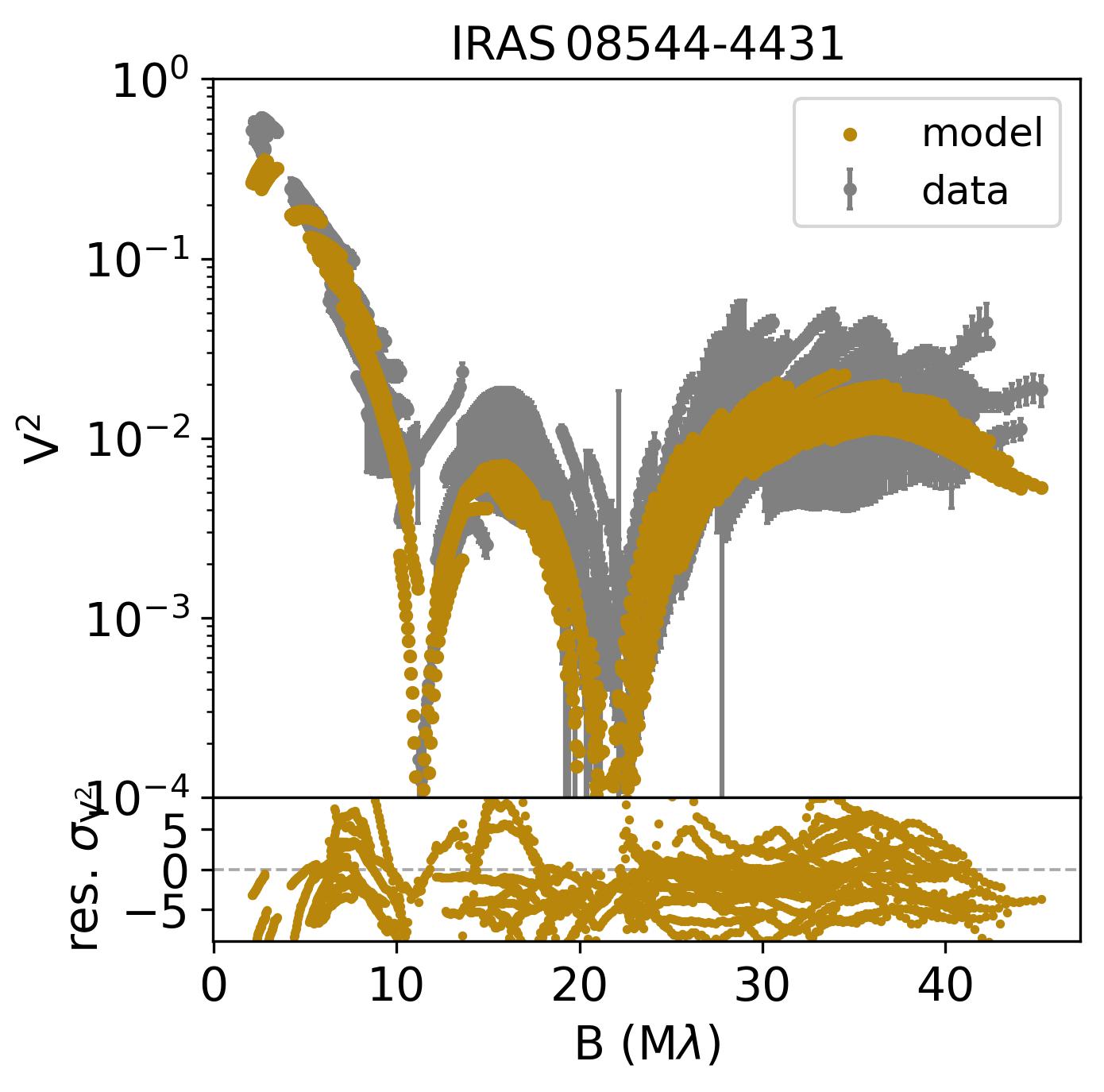}
  \end{minipage}

\begin{minipage}{0.24\textwidth}
  \includegraphics[width=\textwidth,width=1.0
  \textwidth]{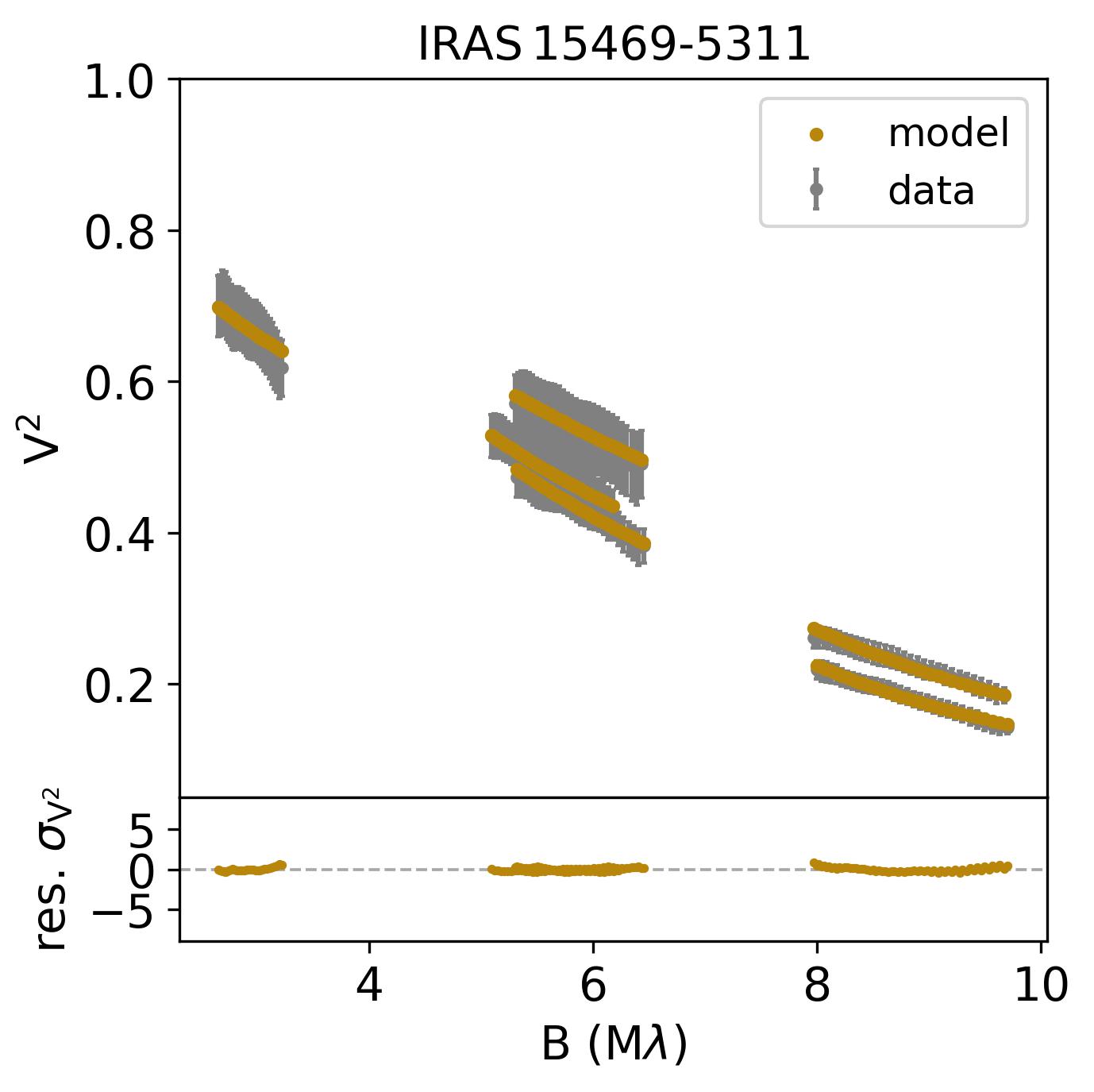}
  \end{minipage}
    \begin{minipage}{0.24\textwidth}
  \includegraphics[width=\textwidth,width=1.0
  \textwidth]{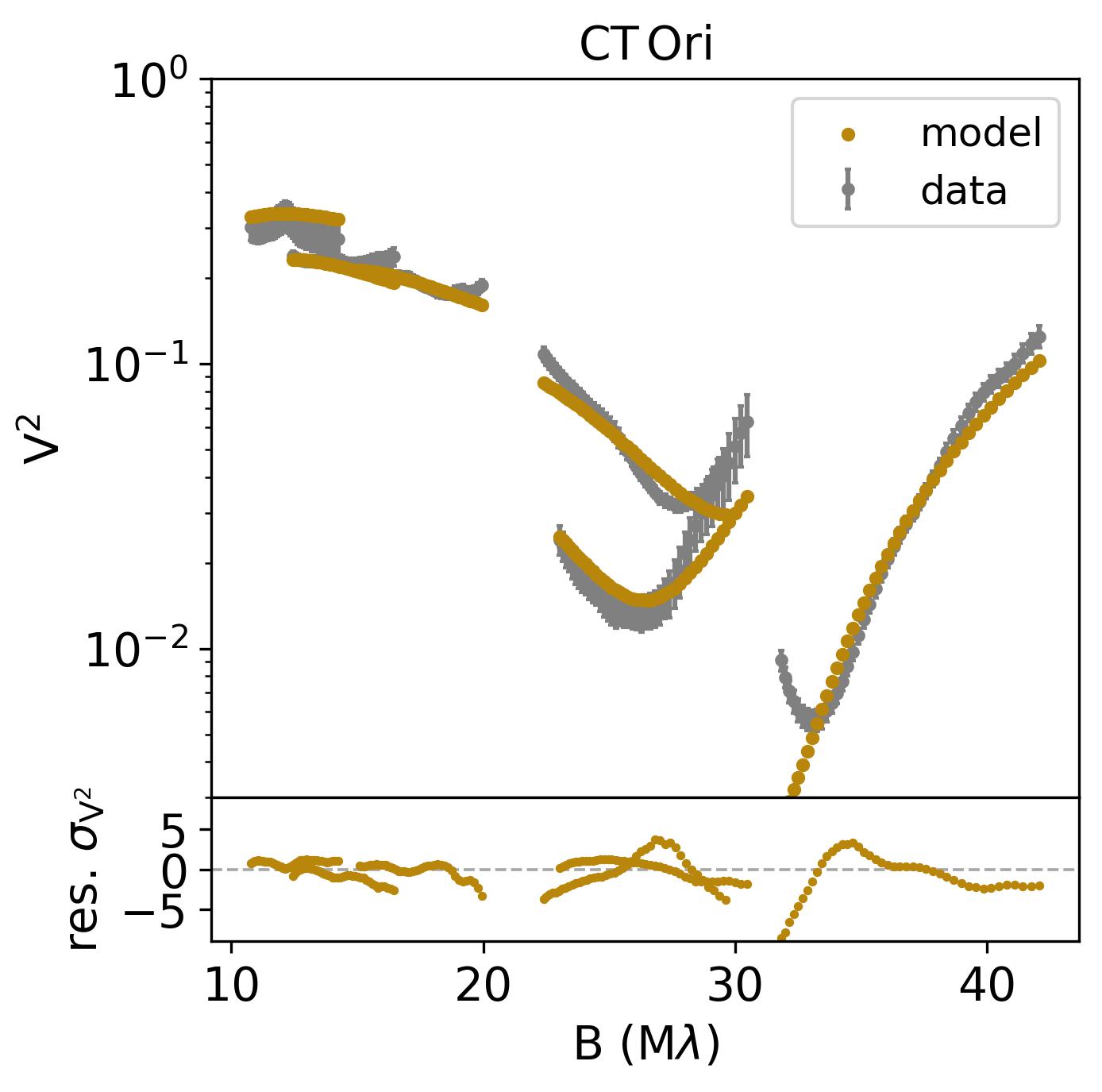}
  \end{minipage}
  \begin{minipage}{0.24\textwidth}
  \includegraphics[width=\textwidth,width=1.0
  \textwidth]{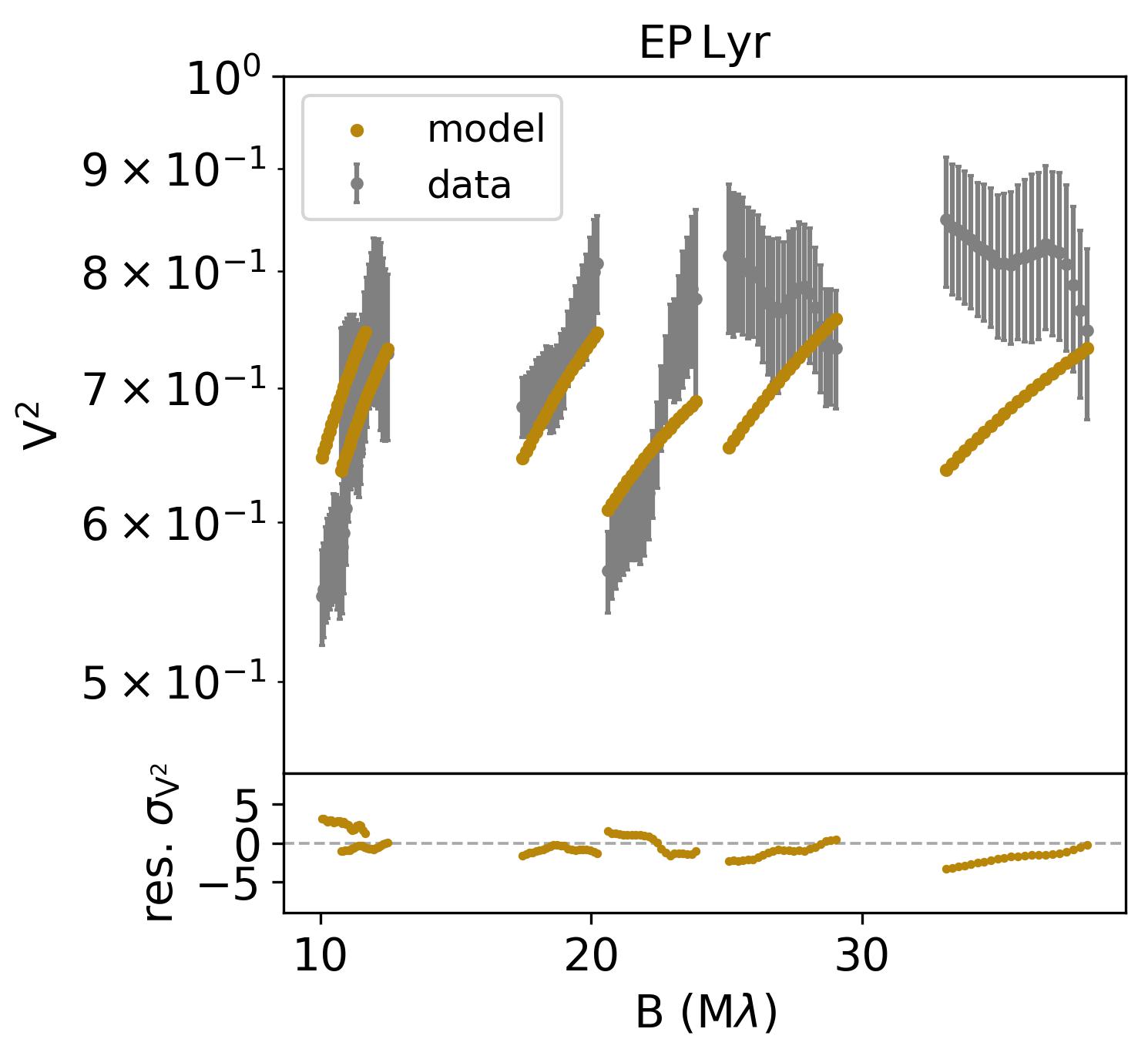}
  \end{minipage}
      \begin{minipage}{0.24\textwidth}
  \includegraphics[width=\textwidth,width=1.0
  \textwidth]{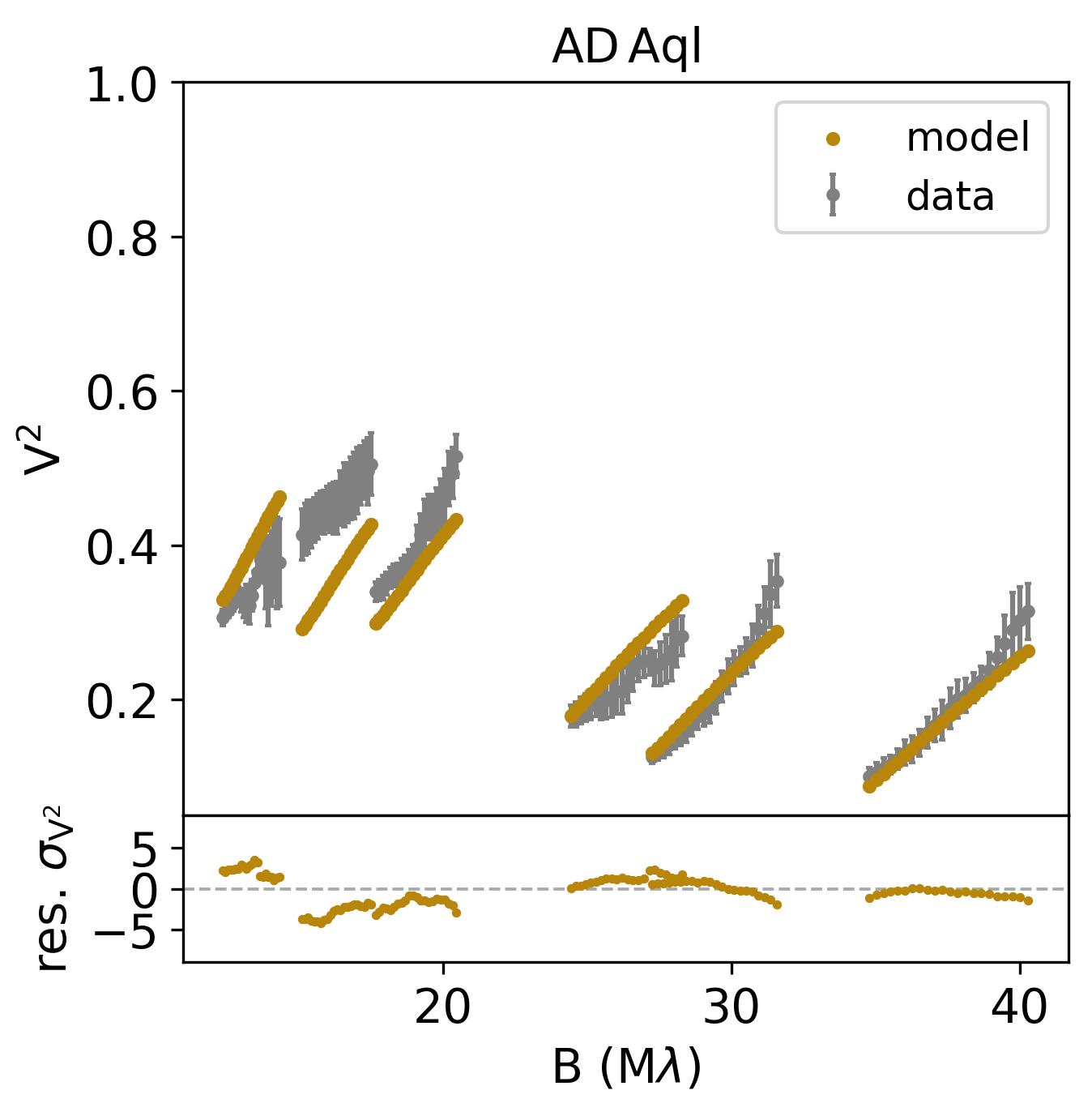}
  \end{minipage}
    \begin{minipage}{0.24\textwidth}
  \includegraphics[width=\textwidth,width=1.0
  \textwidth]{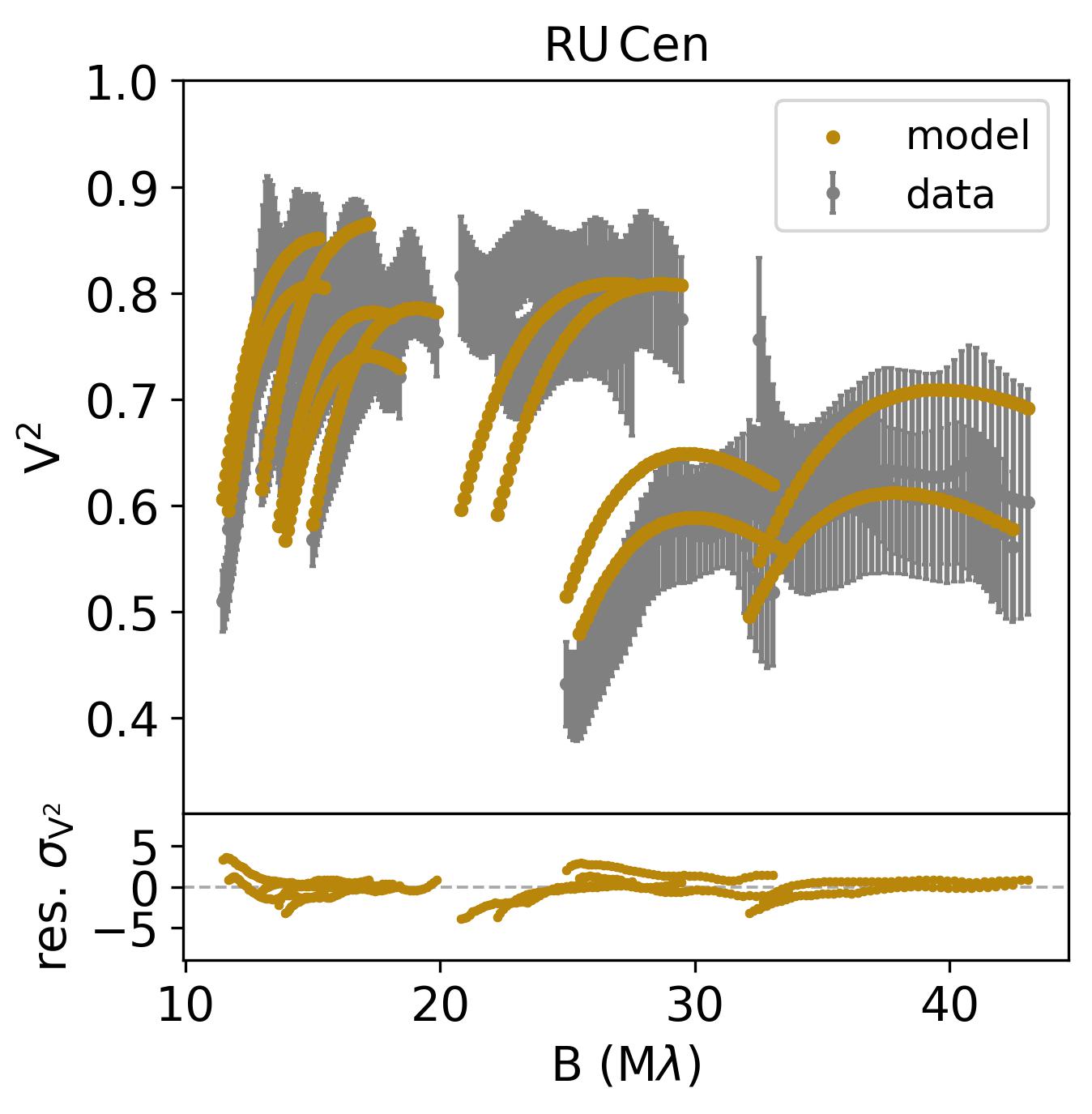}
  \end{minipage}
  \begin{minipage}{0.24\textwidth}
  \includegraphics[width=\textwidth,width=1.0
  \textwidth]{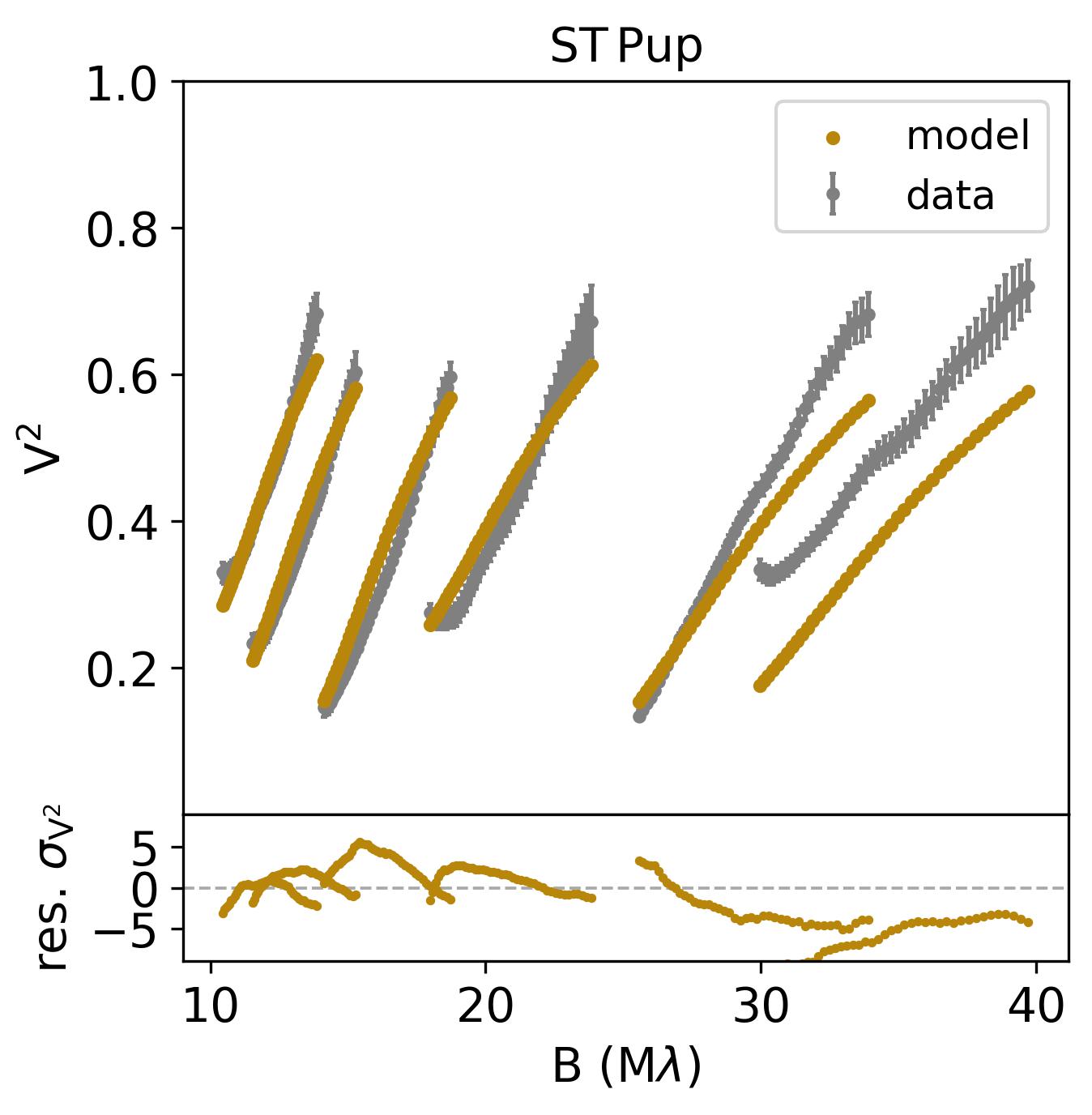}
  \end{minipage}

  \caption{Best-fitting models to the squared visibilities in the $L$ band vs the data of the different targets. 
  The bottom panels show the residuals in $\sigma$.}
  \label{fig:model_fits_Lband}
\end{figure*}

\begin{figure*}

\centering
\begin{minipage}{0.24\textwidth}
  \includegraphics[width=\textwidth,width=1.0
  \textwidth]{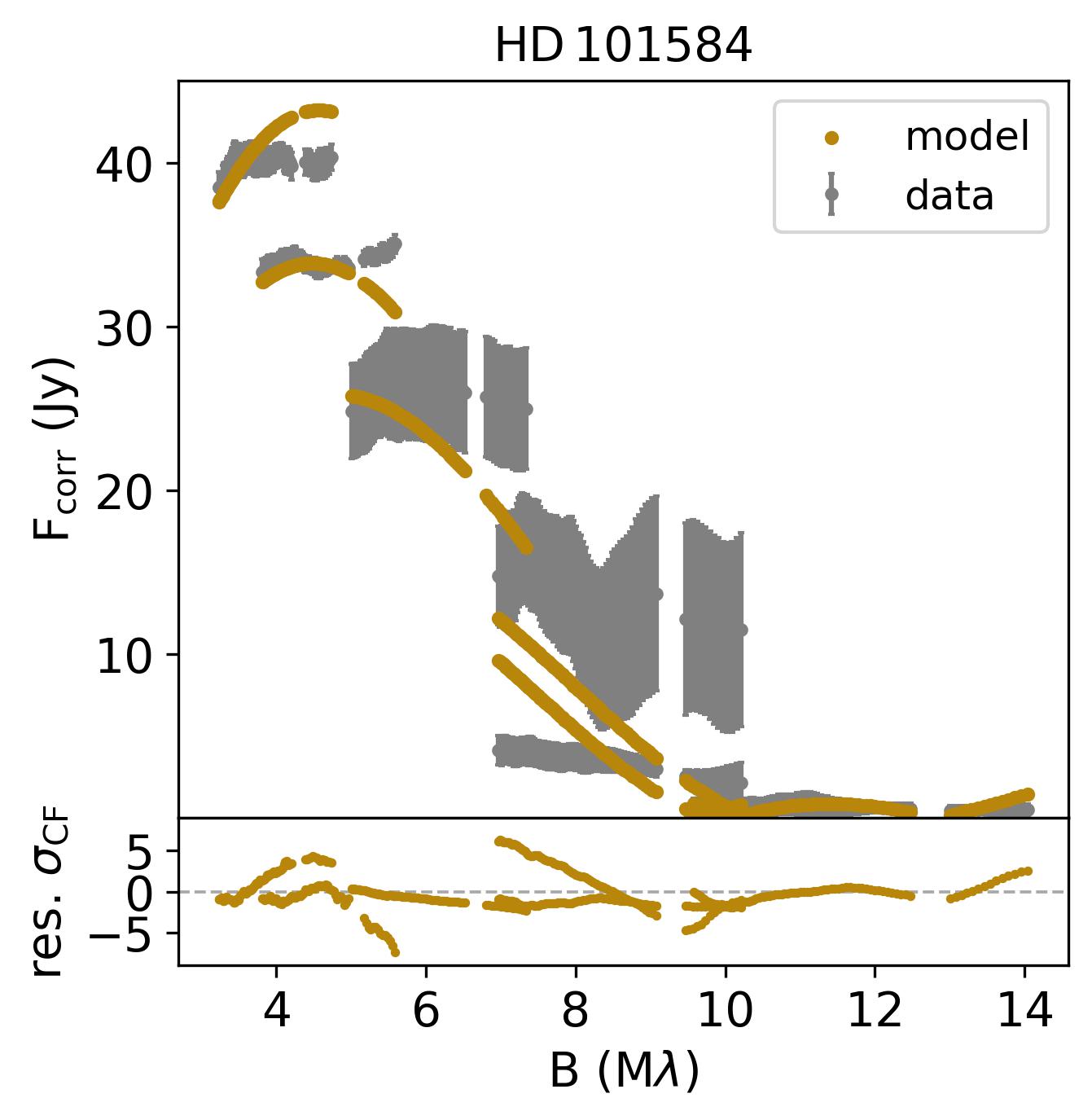}
  \end{minipage}
\begin{minipage}{0.24\textwidth}
  \includegraphics[width=\textwidth,width=1.0
  \textwidth]{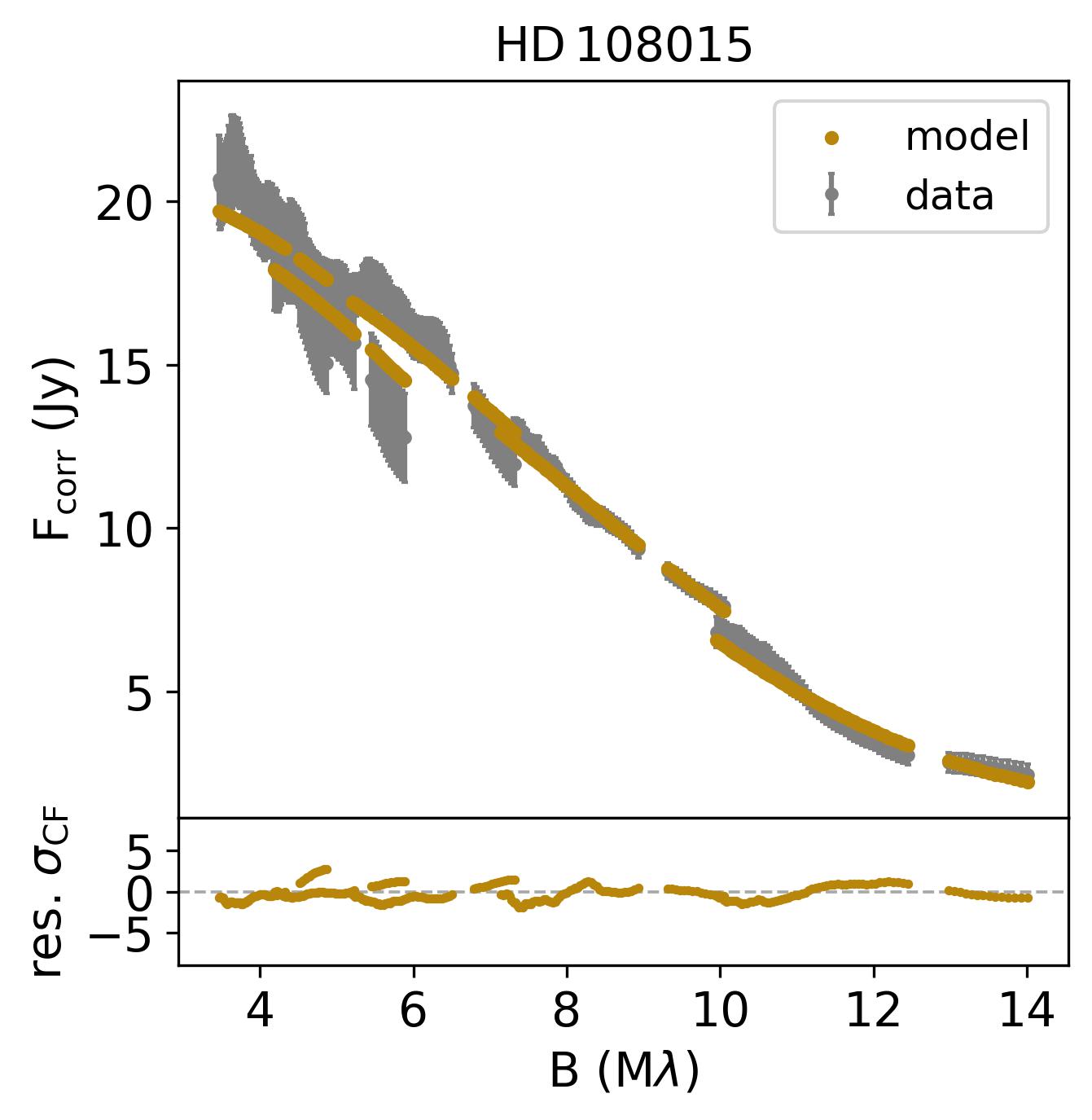}
  \end{minipage}
  \begin{minipage}{0.24\textwidth}
  \includegraphics[width=\textwidth,width=1.0
  \textwidth]{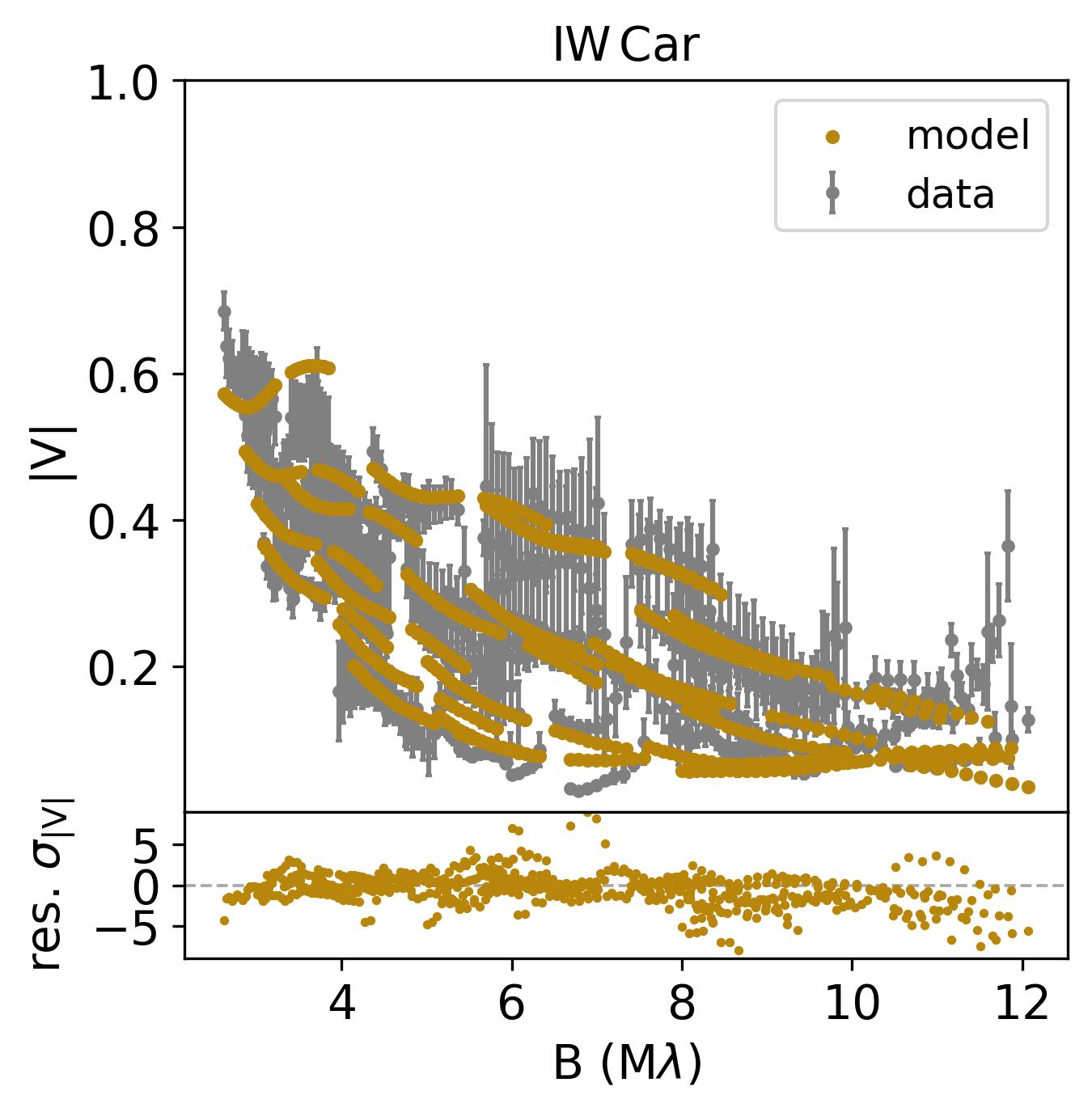}
  \end{minipage}
\begin{minipage}{0.24\textwidth}
  \includegraphics[width=\textwidth,width=1.0
  \textwidth]{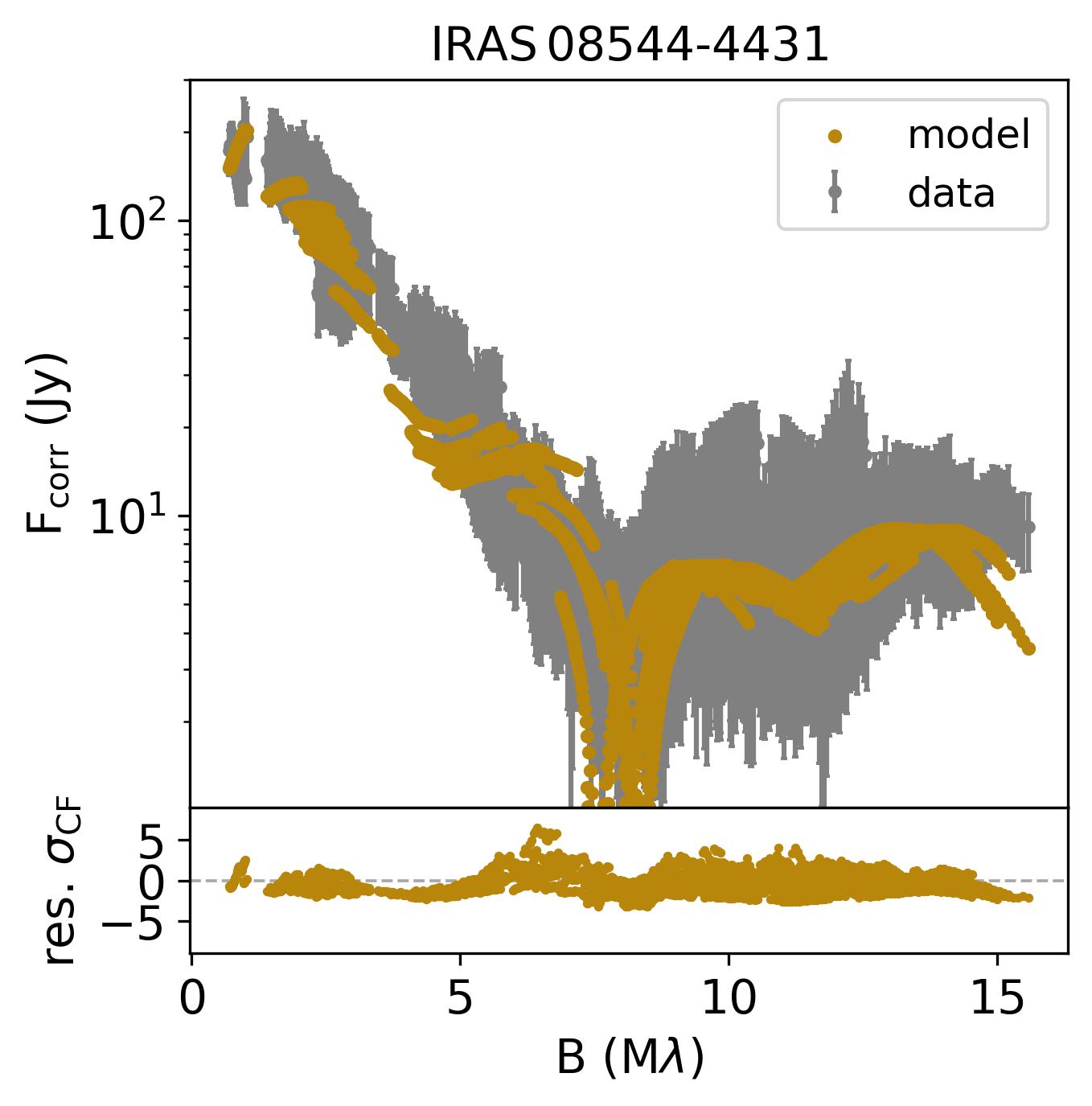}
  \end{minipage}
  \begin{minipage}{0.24\textwidth}
  \includegraphics[width=\textwidth,width=1.0
  \textwidth]{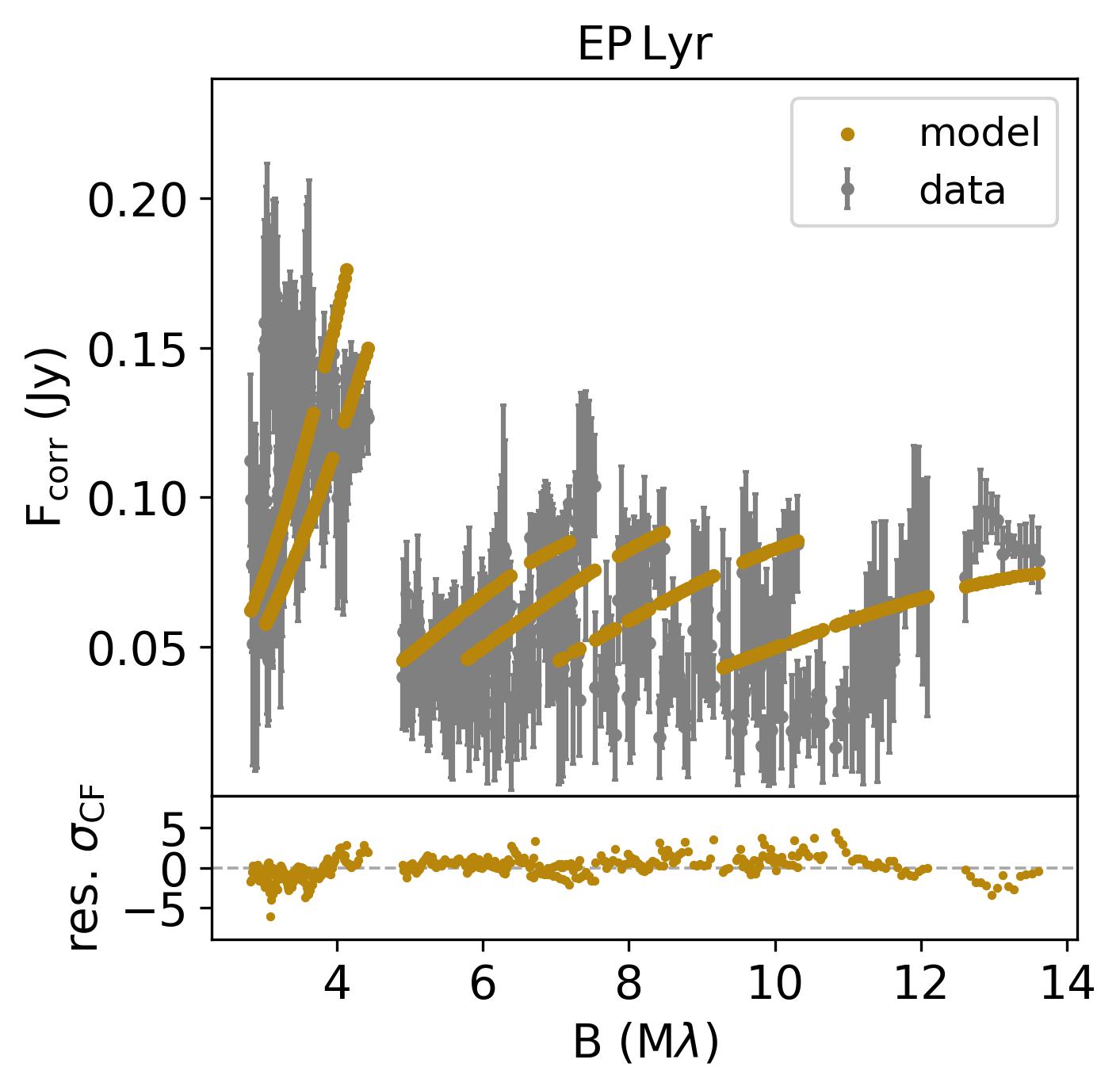}
  \end{minipage}
    \begin{minipage}{0.24\textwidth}
  \includegraphics[width=\textwidth,width=1.0
  \textwidth]{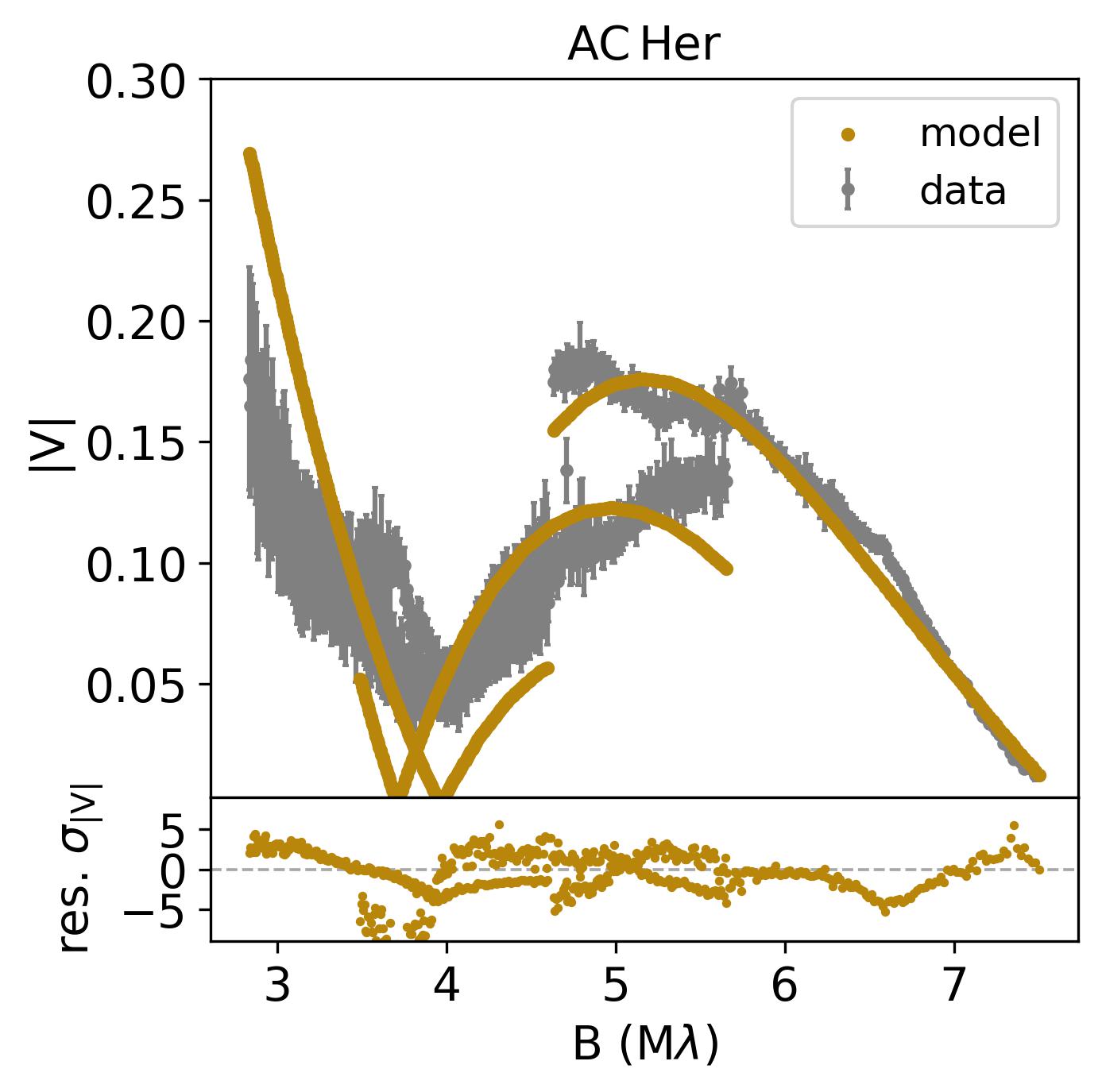}
  \end{minipage}
    \begin{minipage}{0.24\textwidth}
  \includegraphics[width=\textwidth,width=1.0
  \textwidth]{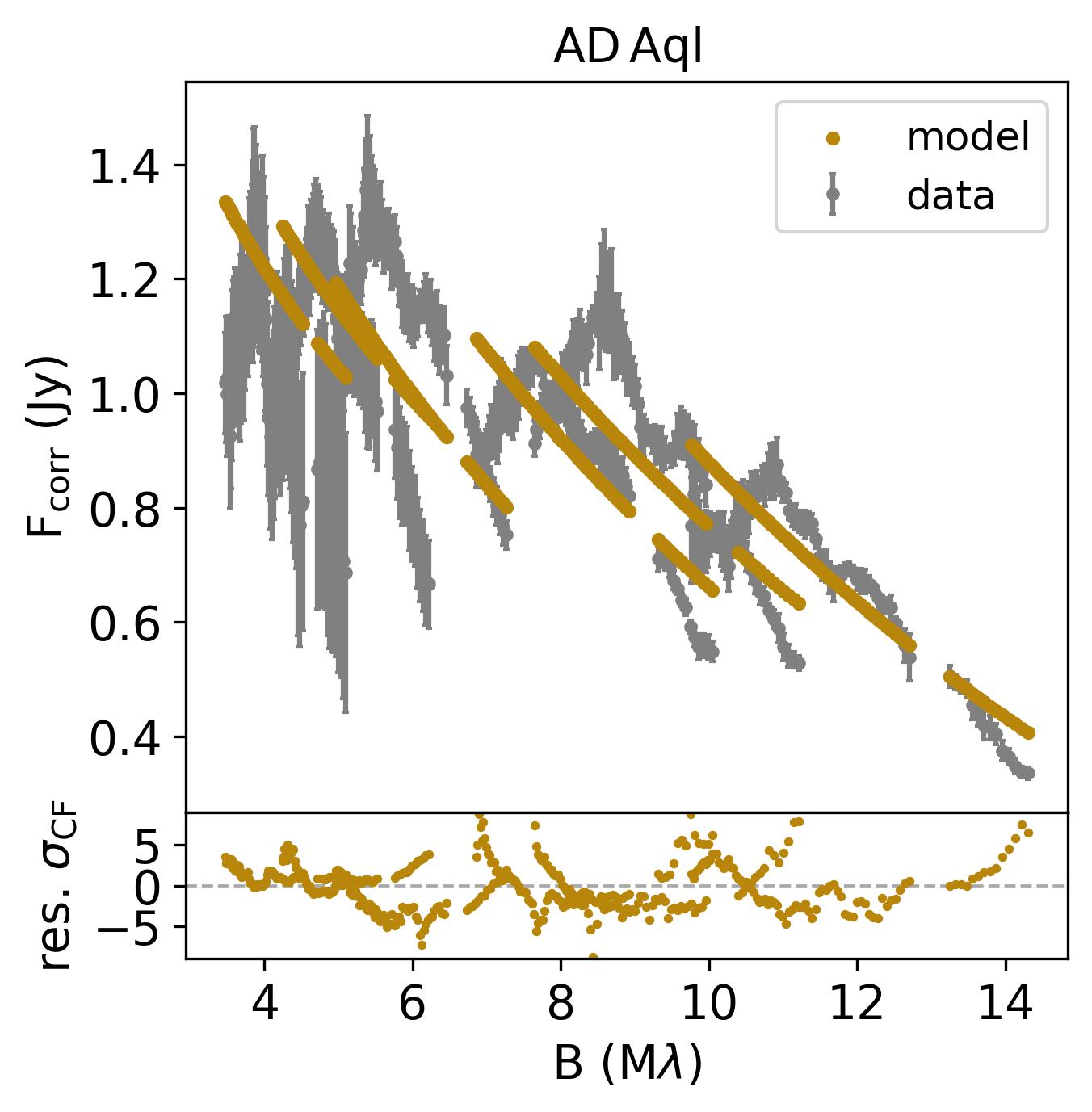}
  \end{minipage}
    \begin{minipage}{0.24\textwidth}
  \includegraphics[width=\textwidth,width=1.0
  \textwidth]{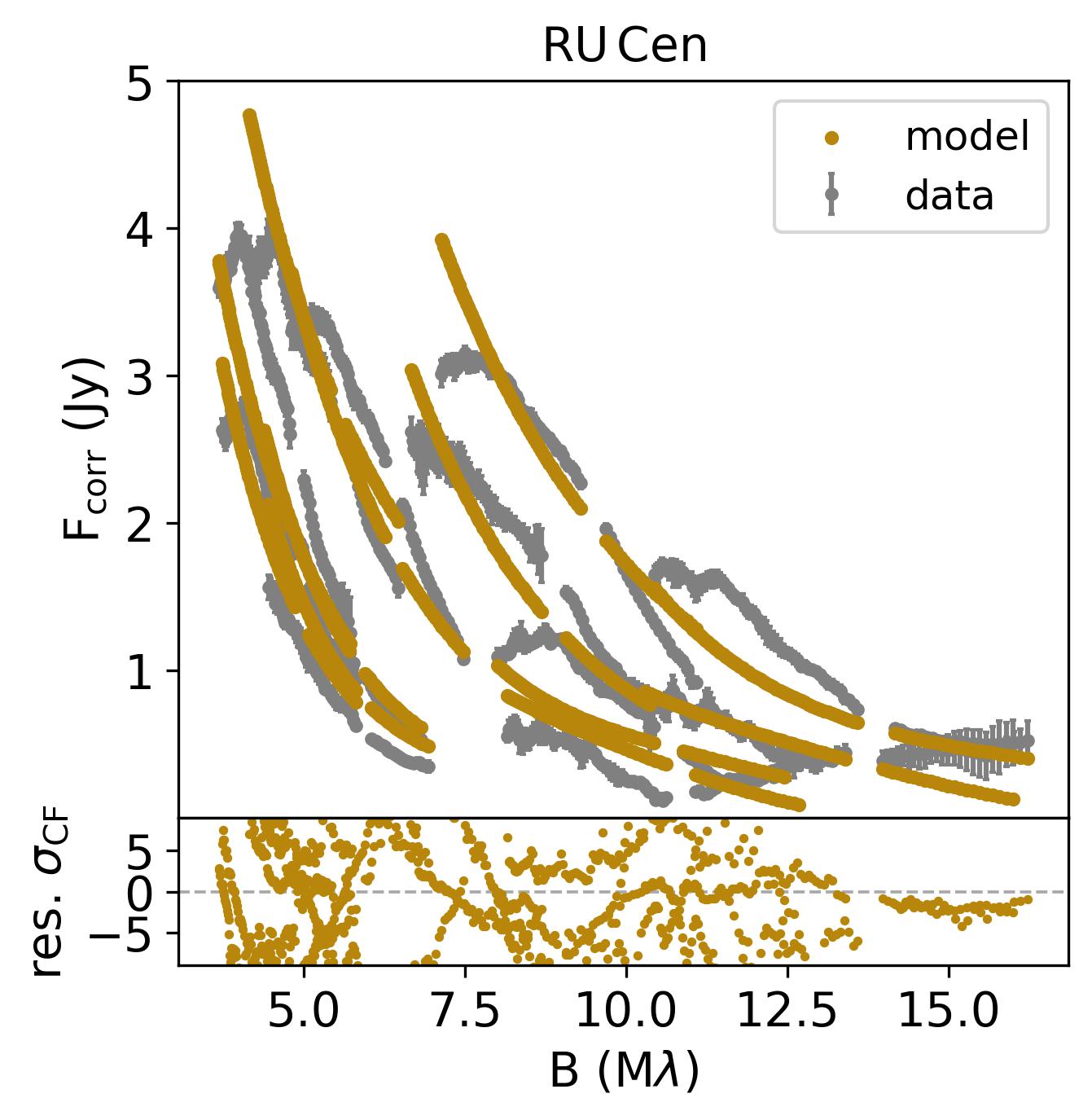}
  \end{minipage}
  \begin{minipage}{0.24\textwidth}
  \includegraphics[width=\textwidth,width=1.0
  \textwidth]{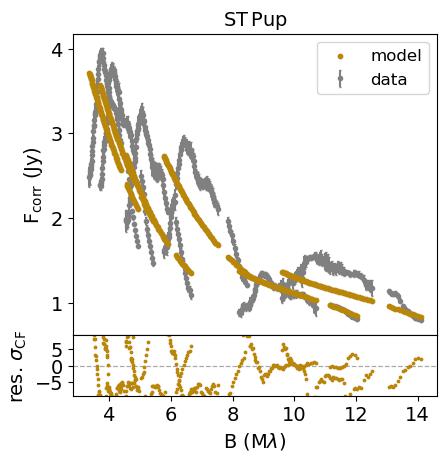}
  \end{minipage}
  \caption{Best-fitting models in the $N$ band vs the data of the different targets. 
  Visibilities are either displayed in correlated fluxes in units of Jansky or in visibility amplitudes (for IW\,Car and AC\,Her).
  The bottom panels show the residuals in $\sigma$.}
  \label{fig:model_fits_Nband}
\end{figure*}

The best-fitting geometric models to the visibilities of the $L$ - and $N$-band data of the 11 targets in our sample are shown in Figs. \ref{fig:model_fits_Lband} and \ref{fig:model_fits_Nband}, respectively.

\end{appendix}
%\endgroup
\end{document}